%% file: CIKM-2020-Tran.tex
\documentclass[sigconf]{acmart}
\AtBeginDocument{%
  \providecommand\BibTeX{{%
    \normalfont B\kern-0.5em{\scshape i\kern-0.25em b}\kern-0.8em\TeX}}}

\usepackage{bm}
\usepackage{amsthm}
\usepackage{array}
\usepackage{algorithm}
\usepackage{algpseudocode}
\usepackage{amsmath}
\usepackage{amssymb}
\usepackage{balance}
\usepackage{booktabs}
\usepackage{bm}
\usepackage{color}
\usepackage{etex}
\usepackage{graphicx}
\usepackage{multirow}
\usepackage{stmaryrd}
\usepackage{subcaption}
\usepackage{url}
\usepackage{thmtools}
\usepackage{mathtools}
\usepackage{todonotes}
\usepackage{ctable}

 \newcommand{\squishlist}{
	\begin{list}{$\bullet$}
		{ \setlength{\itemsep}{0pt}
			\setlength{\parsep}{3pt}
			\setlength{\topsep}{3pt}
			\setlength{\partopsep}{0pt}
			\setlength{\leftmargin}{1.5em}
			\setlength{\labelwidth}{1em}
			\setlength{\labelsep}{0.5em} } }
	
	\newcommand{\squishlisttwo}{
		\begin{list}{$\bullet$}
			{ \setlength{\itemsep}{0pt}
				\setlength{\parsep}{0pt}
				\setlength{\topsep}{0pt}
				\setlength{\partopsep}{0pt}
				\setlength{\leftmargin}{2em}
				\setlength{\labelwidth}{1.5em}
				\setlength{\labelsep}{0.5em} } }
		
		\newcommand{\squishend}{
	\end{list}  }

\ccsdesc[500]{Information systems~Recommender systems}

\keywords{Long-term and short-term user preferences; Quaternion-based recommenders; Quaternion-based attention; adversarial training.}

\copyrightyear{2020}
\acmYear{2020}
\setcopyright{acmcopyright}\acmConference[CIKM '20]{Proceedings of the 29th ACM International Conference on Information and Knowledge Management}{October 19--23, 2020}{Virtual Event, Ireland}
\acmBooktitle{Proceedings of the 29th ACM International Conference on Information and Knowledge Management (CIKM '20), October 19--23, 2020, Virtual Event, Ireland}
\acmPrice{15.00}
\acmDOI{10.1145/3340531.3411926}
\acmISBN{978-1-4503-6859-9/20/10}

\begin{document}
\fancyhead{}

\title{Quaternion-Based Self-Attentive Long Short-term User Preference Encoding for Recommendation}

\author{Thanh Tran}
\authornote{Denotes equal contribution.}
% \vspace{0.1in}
\affiliation{
% \department{Department of Computer Science}
\institution{Worcester Polytechnic Institute}
}
\email{tdtran@wpi.edu}

\author{Di You}
\authornotemark[1]
% \authornote{Denotes equal contribution.}
% \vspace{0.1in}
\affiliation{
% \department{Department of Computer Science}
\institution{Worcester Polytechnic Institute}
}
\email{dyou@wpi.edu}

\author{Kyumin Lee}
% \vspace{0.1in}
\affiliation{
% \department{Department of Computer Science}
\institution{Worcester Polytechnic Institute}
}
\email{kmlee@wpi.edu}


% \affiliation{%


\begin{abstract}
%Recent works in Natural Language Processing has shown distinctive advantages of Quaternion-based representations (i.e. hypercomplex representations) over the traditional real-value representations: (i) latent inter- and intra-dependencies between multidimensional input features are competently captured with the Hamilton product, leading to a more compact interaction representations of input features; (ii) Quaternion-based transformation requires smaller degree of freedoms, encouraging a smaller number of parameters to learn in the neural models. Unfortunately, most of the current recommender systems rely on real-value representations to model users-items relationship.
Quaternion space has brought several benefits over the traditional Euclidean space: Quaternions (i) consist of a real and three imaginary components, encouraging richer representations; (ii) utilize Hamilton product which better encodes the inter-latent interactions across multiple Quaternion components; and (iii) result in a model with smaller degrees of freedom and less prone to overfitting. Unfortunately, most of the current recommender systems rely on real-valued representations in Euclidean space to model either user's long-term or short-term interests. In this paper, we fully utilize Quaternion space to model both user's long-term and short-term preferences. We first propose a QUaternion-based self-Attentive Long term user Encoding (\emph{QUALE}) to study the user's long-term intents. Then, we propose a QUaternion-based self-Attentive Short term user Encoding (\emph{QUASE}) to learn the user's short-term interests. To enhance our models' capability, we propose to fuse \emph{QUALE} and \emph{QUASE} into one model, namely \emph{QUALSE}, by using a Quaternion-based gating mechanism. We further develop Quaternion-based Adversarial learning along with the Bayesian Personalized Ranking (\emph{QABPR}) to improve our model's robustness. 
Extensive experiments on six real-world datasets show that our fused QUALSE model outperformed 11 state-of-the-art baselines, improving 8.43\% at \emph{HIT@1} and 10.27\% at \emph{NDCG@1} on average compared with the best baseline.
% Experiments show the effectiveness of our proposed models over 11 strong state-of-the-art baselines on six real-world datasets. 
%In particular, \emph{QUALSE} (i.e. our fused model) training with \emph{QABPR} loss improves 8.43\% at \emph{HIT@1} and 10.27\% at \emph{NDCG@1} on average compared with the best baseline over the six datasets.

%In this paper, we move beyond the traditional real-value representations and introduce a hypercomplex representations to better model the users-items interactions. In particularly, we not only use Quaternion embeddings to represent for both users and items in the system, but also fully adapt Quaternion representations in transformation layers. Since the users' preferences are depicted in both long-term and short-term scenarios, we first proposed a QUaternion-based self-Attentive Long term user Encoding (\emph{QUALE}) to model the users' long-term interest. Then, we proposed a QUaternion-based self-Attentive Short term user Encoding (\emph{QUASE}) to model the users' short-term interest. To enhance the our models' capability, we proposed to fuse \emph{QUALE} and \emph{QUASE} into one model using a Quaternion-based gated mechanism. For further improving the model's robustness, we proposed a Quaternion-based adversarial learning goes along with the Bayesian Personalized Ranking. Experiments on several datasets has shown the effectiveness of our proposals over traditional real-value recommender systems.
\end{abstract}

\maketitle

\input{1-intro.tex}
\input{2-relatedwork.tex}
\input{3-preliminary.tex}
\input{4-model.tex}
\input{5-experiments.tex}
\input{6-conclusion.tex}

\small
\bibliographystyle{ACM-Reference-Format}
\bibliography{ref}

\end{document}

%% file: 1-intro.tex
%\vspace{-5pt}
\section{Introduction}
\label{sec:intro}
% In the era of information explosion, 
Recommender Systems \cite{resnick1997recommender} have become the heart of many online applications such as e-commerce, music/video streaming services, social media, \emph{etc}. Recommender systems proactively helped (i) users to explore new/unseen items, (ii) potentially the users stay longer on the applications, and (iii) companies increase their revenue.
%Recent advances in recommendation systems have divided them into two main streams: (i) general recommendation, and (ii) sequential recommendation.

%talk about general recommendation and conclude they are limited because of only long-term preference modeling.
Matrix Factorization techniques \cite{koren2009matrix,hu2008collaborative,he2016fast} extracted features of users and items to compute their similarity. Recently, deep neural network boosted performance of a recommender system by providing non-linearity which helped modeling complex relationships between users and items \cite{he2017neural}. However, these prior works only focused on a user and a target item without considering the user's previously consumed items, some of which may be related to the target item.
%A limitation for this line of work is that, in real world scenarios, users' history behaviors usually have a bearing on his/her next consumption.
While some prior works \cite{koren2008factorization,kabbur2013fism} largely premised on unordered user interactions, users' interests are intrinsically dynamic and evolving. Based on the observation, 
% other existing works
\cite{he2017translation,rendle2010factorizing,tang2018personalized,kang2018self,hidasi2015session} followed two paradigms to capture a user's sequential pattern: (i) \emph{short-term} item-item transitions, or (ii) \emph{long-term} item-item transitions. %The most well-known methods to model \emph{short-term} item-item transitions include the adoption of first order Markov chain to explore the transitions from the most recently consumed item to the next item \cite{he2017translation,rendle2010factorizing}, or using neural network architectures to model item-item transitions with a higher order Markov chain \cite{tang2018personalized}. For modeling \emph{long-term} item-item transitions, existing works adapted self-attention in Transformer network \cite{kang2018self}, or exploited RNN-based architectures \cite{hidasi2015session}.

%\begin{figure}[t]
%    \centering
%    \includegraphics[width=0.86\linewidth]{figs/motivation_example.pdf}
%    \vspace{-5pt}
%    \caption{A toy example illustrating the existence of both long-term and short-term item dependencies.}
%    \label{fig:motivation_example}
%    \vspace{-15pt}
%\end{figure}

\begin{figure}[t]
    \centering
    \begin{subfigure}{0.23\textwidth}
        \centering
        \includegraphics[width=0.98\textwidth, height=80pt]{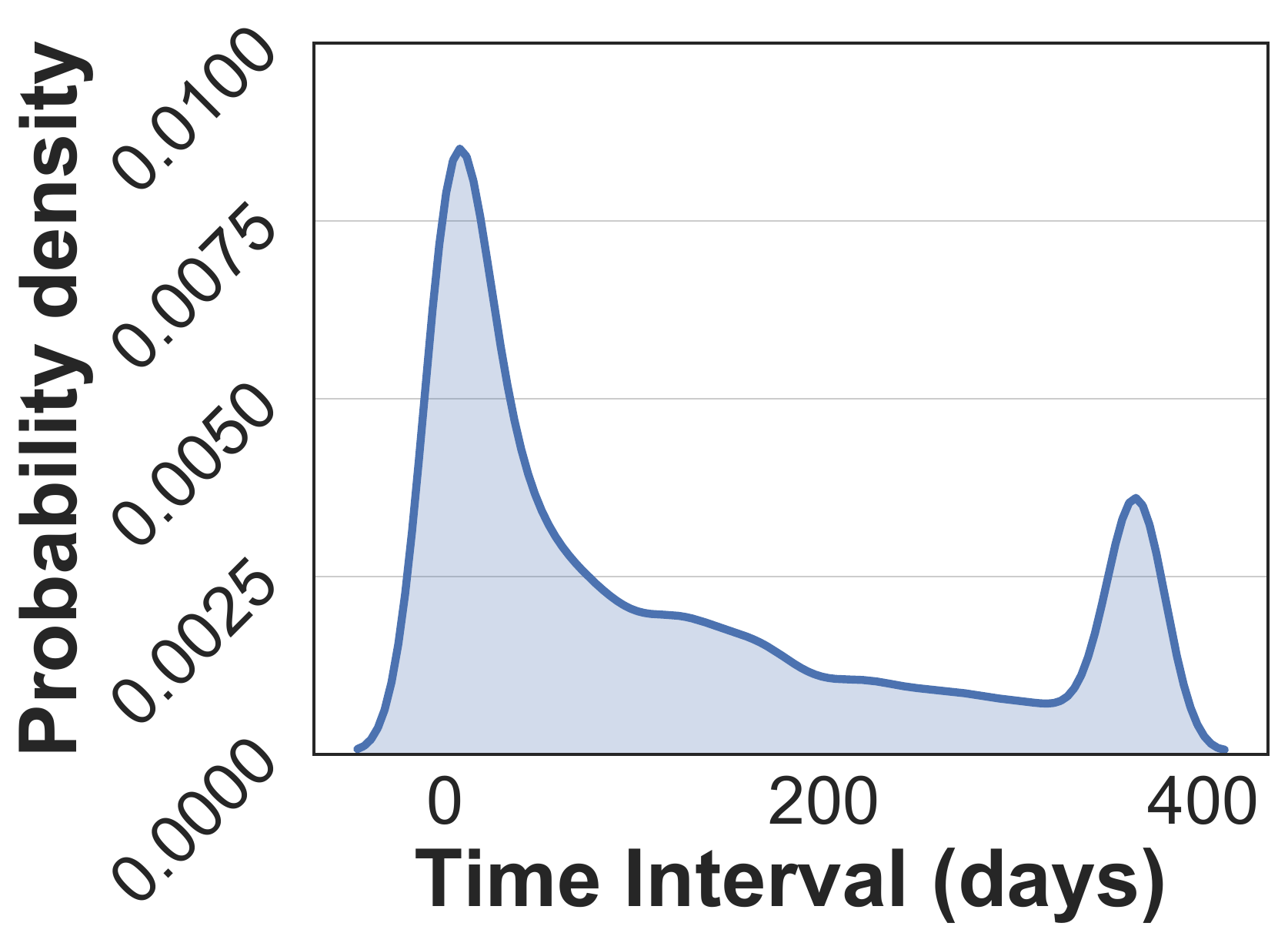}
        \vspace{-5pt}
        \caption{Video Games}
        \label{fig:video_train_density}
    \end{subfigure}
    \begin{subfigure}{0.23\textwidth}
        \centering
        \includegraphics[width=0.98\textwidth,height=80pt]{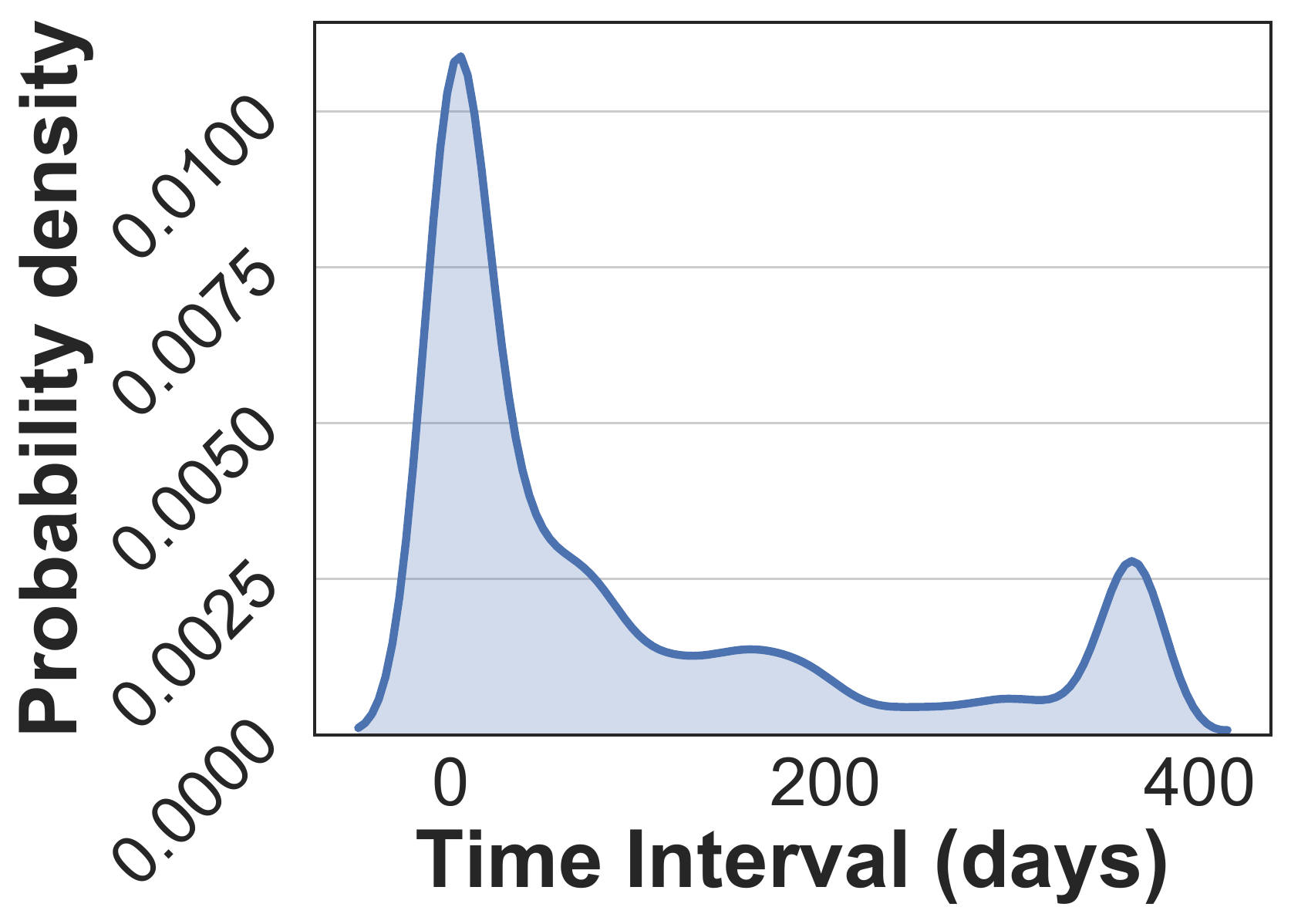}
        \vspace{-5pt}
        \caption{Toys and Games}
        \label{fig:toys_train_density}
    \end{subfigure}
    \vspace{-10pt}
    \caption{Density distribution of item-item similarity scores on Amazon \emph{Video Games}, and \emph{Toys and Games} datasets.}
    \label{fig:train_stat}
    \vspace{-4ex}
\end{figure}

\begin{figure*}[t]
    \centering
    \includegraphics[width=0.71\textwidth]{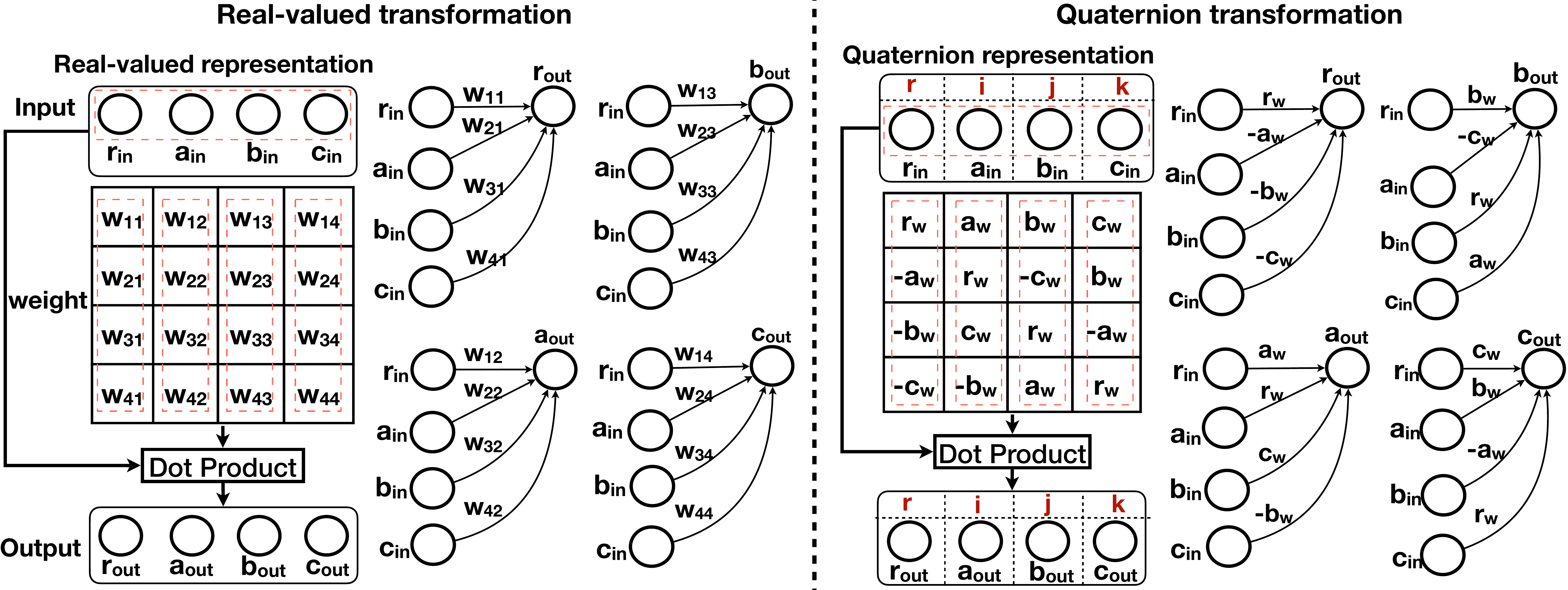}
    \vspace{-10pt}
    \caption{Comparison between real-valued transformation (Left) and Quaternion transformation (Right). We replace Hamilton product in Quaternion space with an equivalent dot product in real space for an easy reference. %Calculating each output dimension in real-valued transformation always need 4 new parameters, resulting in 16 degrees of freedom. In contrast, only 4 parameters are used and shared in producing all output dimensions in Quaternion transformation, leading to a better inter-dependencies encoding and a smaller degrees of freedom requirement.
    }
    \vspace{-10pt}
    \label{fig:motivation}
\end{figure*}

However, user's interests can be highly diverse, so modeling only either \emph{short-term} or \emph{long-term} user intent does not fully capture the user's preferences, producing less effective recommendation results.
%To illustrate the point, we describe a toy example, where a user watched two movies ``Avengers 2'' and ``Ghost in the Shell'' long time ago (i.e., long-term user interest), and recently watched three movies ``Captain Marvel'', 'Spiderman', and ``Glass'' (i.e., short-term user interest). Given the target movie ``Avengers 4'', we can see the target movie depends on both short-term and long-term consumed items/interests. We also conducted empirical analysis on \emph{Amazon} \emph{Video Games}, and \emph{Toys and Games} datasets to verify the hypothesis (see Figure \ref{fig:train_stat}).
To illustrate the point, we conducted an empirical analysis on \emph{Amazon} \emph{Video Games}, and \emph{Toys and Games} datasets. First, we represent each item by a multi-hot encoding, where item $j$ is represented by a vector $t \in \mathbb{R}^m$, position $i=1$ if user $i$ consumed the current item, and $m$ denotes the total number of users in a dataset. For each user, her consumed items are sorted in the chronological order. Then, we calculated a cosine similarity score between each item and each of its previously consumed items. Then we selected the largest cosine similarity score per item per user. Figure \ref{fig:train_stat} presents the density distribution of the consumed time interval (\emph{x-axis}) between each pair of item and its most similar previously consumed item. We observe that there exists a bimodal distribution, where one (\emph{left}) peak lays at a relative short-term period and the other (\emph{right}) peak locates in a long-term period. The observation confirms that both long-term and short-term preferences played important roles on the user's current purchasing intent. We observe the same phenomenon from the other four datasets described in Section~\ref{sec:exp}. %Note that other datasets admit the same observations but are omitted due to the space limitation.
%We first adopt the multi-hot encoding for each item, where item $j$ is represented by a vector $t \in \mathbb{R}^m$ where position $i=1$ if user $i$ consumed current item. Such encoding method is simple yet enrich the item representation with historical interaction information between users and items. For each user behavior sequence consists of numerous chronically ordered items, we calculate the cosine similarity score between each item and all its previous items. Intuitively, the most similar previous item should have the highest score. Figure\ref{fig:train_stat} demonstrate the density distribution of time interval between each item and its most similar previous item. It is obvious that there is a bimodal distribution, where one peak lies at relative short period and another locates in long-term. The observation confirms that both long-term and short-term preferences have an effect on current purchasing intent.

%\noindent\textbf{Motivation:} 
Based on the observation, we propose a Quaternion-based neural recommender system that models both long-term and short-term user preferences. Unlike the prior works \cite{zhou2019deep,yu2019adaptive} which rely on Euclidean space, our proposed recommender system models both user's long-term and short-term preferences in a hypercomplex system (i.e., Quaternion Space) to further improve the recommendation quality.
%Though some recent works modeled both \emph{long-term} and \emph{short-term} user intents \cite{zhou2019deep,yu2019adaptive}, existing recommender systems mostly relied on Euclidean space. In these models, users, items and neural transformations -- the heart of neural networks -- were characterized by real-valued representations. Unlike the prior works, we explore a new way to design a recommender system that models both long-term and short-term user preferences in a hypercomplex system (i.e., Quaternion Space) 
Concretely, we utilize Quaternion representations for all users, items and neural transformations in our proposed models. There are numerous benefits of the Quaternion utilization over the traditional real-valued representations in Euclidean space: (1) Quaternion numbers/vectors consist of a real component and three imaginary components (i.e. $\boldsymbol{i, j, k}$), encouraging a richer extent of expressiveness; (2) instead of using dot product in Euclidean space, Quaternion numbers/vectors operate on Hamilton product, which matches across multiple (inter-latent) Quaternion components and strengthens their inter-latent interactions, leading to a higher expressive model; (3) the weight sharing nature of Hamilton product leads to a model with a smaller number of parameters.

To illustrate these benefits of the Quaternion utilization, we show a comparison of a transformation process with Quaternion representations \emph{vs.} real-valued representations in Figure \ref{fig:motivation}. In Euclidean space, different output dimensions are produced by multiplying the same input with different weights. Given a real-valued 4-dimensional vector $[r_{in}, a_{in}, b_{in}, c_{in}]$, it takes a total of 16 parameters (i.e. 16 degrees of freedom) to transform into $[r_{out}, a_{out}, b_{out},$ $c_{out}]$. For Quaternion transformation, the input vector now is represented with 4 components, where $r_{in}$ is the value of the real component, $a_{in}$, $b_{in}$, $c_{in}$ are the corresponding values of the three imaginary parts $\boldsymbol{i}$, $\boldsymbol{j}$, $\boldsymbol{k}$. Due to the weight sharing nature of Hamilton product (refer to the Equa~(\ref{equa:quaternion-hamiltonprod}) in Section \ref{sec:preliminary}), different output dimensions take different combinations of the same input with only 4 weighting parameters \{$r_{w}, a_{w}, b_{w}, c_{w}$\}. The Quaternions provide a better inter-dependencies interaction coding and reduce 75\% of the number of parameters compared with real-valued representations in Euclidean space (e.g., 4 unique parameters vs. 16 parameters).

%Recently, Quaternion space has shown its effectiveness in speech recognition \cite{parcollet2018quaternion}, computer vision \cite{gaudet2018deep}, and NLP applications \cite{tay2019lightweight}.
To our best of knowledge, we are the first work that fully utilizes Quaternion space in modeling both user's long-term and short term interests. Furthermore, to increase our model's robustness, we propose a Quaternion-based Adversarial attack on Bayesian Personalized Ranking (QABPR) loss. As far as we know, we are the first, applying adversarial attack on Quaternion representations in the recommendation domain.

We summarize our contributions in the paper as follows:
%\begin{itemize}
\squishlist
    \item We propose novel Quaternion based models to learn a user's long-term and short-term interests more effectively. As a part of our framework, we propose Quaternion self-attention that works in Quaternion space.
    \item We propose a Quaternion-based Adversarial attack on BPR-loss to further improve the robustness of our model.
    \item We conduct extensive experiments to demonstrate the effectiveness of our models against 11 strong state-of-the-art baselines on six real-world datasets.
    % \item We propose to study both user dynamic short-term intent and global long-term interest using our Quaternion-based neural networks. To the best of our knowledge, we are the first to fully utilize Quaternion space in recommendation domain.
    % \item We propose Quaternion based adversarial attack on Bayesian Personalized Ranking loss to enhance our models' robustness. To the best of our knowledge, we are the first work proposing the adversarial attack on Quaternion space.
\squishend

%% file: 2-relatedwork.tex
\vspace{-5pt}
\section{Related Work}
%Recommendation systems using user-item interactions are mostly divided into two groups: (i) general recommenders, and (ii) sequential recommenders. General recommenders exploited all user-item interactions without considering the order of consumed items. Sequential recommenders took into account consumed items in order. %We summarize and relate them to our work as follows:

%In our context of modeling both long and short-term user interests, for an easy reference, we re-categorized them into two groups: (i) \emph{Group 1}: long-term user preference modeling and (ii) \emph{Group 2}: short-term user preference modeling. Note that general recommenders are belong to \emph{Group 1} while some of sequential recommenders belong to \emph{Group 1} and others belong to \emph{Group 2}, depending on how ``\emph{long-term}'' items are considered. We summarize them as follows:

%\smallskip
\noindent\textbf{General Recommenders.}
%\subsection{General Recommenders}
Matrix Factorization is the most popular method to encode global user representations by using unordered user-item interactions \cite{hu2008collaborative,koren2009matrix,zhang2016discrete}. Its basic idea is to represent users and items by latent factors and use dot product to learn the user-item affinity.
Despite their success, they cannot model non-linear user-item relationships due to the linear nature of dot product. To overcome the limitation, neural network based recommenders were recently introduced \cite{he2017neural,vinh2019interact,tran2019adversarial,ebesu2018collaborative}. 
\cite{he2017neural} combined a \emph{generalized matrix factorization} component and a non-linear user-item interactions via a MLP architecture. 
% \cite{he2017neural} proposed a neural recommender that learned a \emph{generalized matrix factorization} component, and non-linear user-item interactions via a MLP architecture. 
\cite{ma2018point,liang2018variational,ma2019gated} substituted the MLP architecture with the auto-encoder design. \cite{xin2019relational,tran2019signed} used memory augmentation to learn different user-item latent relationship. 
% However, these methods are still limited by only considering the user-item latent space, and overlooked the correlations in the item-item latent space. Moreover, 
When non-existed users come with some observed interactions (i.e., recently created user accounts with some item interactions), the recommenders need to be rebuilt to generate their representations.
To avoid these issues, current works encode users by combining the users' consumed item embeddings in two main streams: (i) taking average of the consumed items' latent representations \cite{koren2008factorization,kabbur2013fism}, or (ii) attentively summing \cite{he2018nais} the consumed items' embeddings.

Despite their success, general recommenders mostly consider all users' unordered consumed items, and produce global/long-term user representations, which are supposed to be static, or changed slowly. Thus, they failed to capture the user's dynamic behavior, that is captured by the user's short-term preference 
% (refer to Figure \ref{fig:train_stat} for illustration).
(see Figure \ref{fig:train_stat}).

%Here we review several classes of previous work related to sequential recommendation as well as quaternion-based modeling.
% \vspace{-5pt}
\smallskip
\noindent\textbf{Sequential Recommenders.}
%\subsection{Sequential Recommenders}
%Due to the highly practical value, recommender systems have attracted increasing research interests and many methods have been developed to address observed problems.
Sequential recommendation is known for its superiority to capture temporal dependencies between historical items \cite{ma2019hierarchical}. Early works relied on Markov Chains to capture item-item sequential patterns \cite{feng2015personalized,rendle2010factorizing,he2017translation}. Other works exploited the convolution architecture to capture more complex temporal dependencies \cite{tang2018personalized}. These methods used short-term item dependencies to model a user's dynamic interest.
Other sequential recommenders focused on modeling long-term user preferences using RNN-based architectures  \cite{hidasi2015session,wu2017recurrent,li2017neural,liu2016context}. However, modeling either long-term user interests or short-term user interests is suboptimal since they concurrently affect a user's intent (Figure \ref{fig:train_stat}). Recent works combined both long and short-term user preferences in real-valued representations to obtain satisfactory results \cite{kang2018self,zhou2019deep, yu2019adaptive}.
%Other sequential recommenders focused on modeling long-term user preferences by adopting the multi-head self-attention mechanism in Transformer  \cite{kang2018self}, or the RNN-based architecture \cite{hidasi2015session,wu2017recurrent,li2017neural,liu2016context}. However, modeling either long-term user interests or short-term user interests is suboptimal since they concurrently affect a user's intent (Figure \ref{fig:train_stat}). Recent works combined both long and short-term user preferences in real-valued representations to obtain satisfactory results \cite{zhou2019deep, yu2019adaptive}.

%In this paper, we also consider to model both long-term and short-term user preferences.
Compared with the prior works 
which used real-valued representations
, we propose Quaternion-based models to capture the user's long-term and short-term interests. Quaternion was first introduced by \cite{hamilton1844lxxviii} and it has recently shown its effectiveness over real-valued representations in NLP and computer vision domains \cite{zhu2018quaternion,gaudet2018deep,parcollet2019quaternion,tay2019lightweight}. 
%We acknowledge that \emph{QCF} model \cite{zhang2019quaternion} is the first work using Quaternion space to solve a recommendation problem. 
We acknowledge that \emph{QCF} model \cite{zhang2019quaternion} is the first Quaternion-based recommender. 
However, the authors used it as a simple extension of the matrix factorization method, where users and items are Quaternion embeddings. Thus, the benefits of Quaternion representation were not fully exploited in their network. Furthermore, they designed the model for a general recommendation problem, which has an inherent limitation of only modeling the user's global interest. %Unlike the prior work, we propose a Quaternion adversarial attack on the Bayesian Personalized Ranking loss to further improve the robustness of our proposed models.

%% file: 3-preliminary.tex
\section{Problem Definition}\label{preliminaries}
Denote U=\{$u_1, u_2, ..., u_m$\} as a set of all users where $m = \vert U \vert$ is the total number of users, and P=\{$p_1, p_2, ..., p_n$\} as a set of all items where $n=\vert P \vert$ is the total number of items. Bold versions of those variables, which we will introduce in the following sections, indicate their respective latent representations/embeddings. Each user $u_i \in U$ consumes items in $P$, denoted by a chronological list $T^{(u_i)}$. We denote $L^{(u_i)}$ as the chronological list of long-term consumed items of $u_i$, and $S^{(u_i)}$ as the chronological list of short-term consumed items of $u_i$ (i.e. $s$ most recently consumed items in chronological order of the user $u_i$), $L^{u_i} \cup S^{u_i} =  T^{(u_i)}$. Note that bold versions of $\boldsymbol{i}, \boldsymbol{j}, \boldsymbol{k}$ are used to indicate the three imaginary parts of a Quaternion, while their subscript versions are used as indices.

In this work, we propose and build Quaternion-based recommender systems by using both long-term and short-term user interests, denoted as $P(p_j| L^{(u_i)}, S^{(u_i)})$. Under an assumption that $L^{(u_i)}$ and $S^{(u_i)}$ are independent given the target item $p_j$, we model $P(p_j| L^{(u_i)}, S^{(u_i)})$ by modeling the user's long-term interest $P(p_j| L^{(u_i)})$ and short-term interest $P(p_j | S^{(u_i)})$ separately by using two different Quaternion-based neural networks. Then, we automatically fuse the two models to build a more effective recommender system.
%Recommending next items for the users using only her short-term consumed items is sub-optimal because the model lacks an ability to connect long-term consumed items with the next items. Also, modeling the next items using long-term consumed items only limits the models to learn dynamic users' interests. Hence, in this paper, we aim to suggest next items for users using both users' long-term and short-term consumed items. More specifically, we build Quaternion-based neural networks to approximate the probability of the next/target item to be consumed by each user $u_i$ in the system as $P(p_j| L^{(u_i)}, S^{(u_i)})$. Under an assumption that $L^{(u_i)}$ and $S^{(u_i)}$ are independent given the target item $p_j$, we model $P(p_j| L^{(u_i)}, S^{(u_i)})$ by modeling $P(p_j| L^{(u_i)}$ and $P(p_j | S^{(u_i)})$ separately using two different Quaternion-based neural networks. Then, we automatically fuse the two models into one.

%We first describe preliminaries on Quaternion in Section~\ref{sec:preliminary}. Then, we present our Quaternion-based models to approximate $P(p_j| L^{(u_i)}$, and $P(p_j | S^{(u_i)})$, and describe our fused model in Section~\ref{sec:models}.

%  \vspace{-5pt}
\section{Preliminary on Quaternion}
\label{sec:preliminary}
In this section, we cover important background on Quaternion Algebra and Quaternion Operators that we use to design our models.

\noindent\textbf{Quaternion number:} %\footnote{https://en.wikipedia.org/wiki/Quaternion}:
In mathematics, Quaternions are a hypercomplex number system. A Quaternion number \emph{X} in a Quaternion space $\mathbb{H}$ is formed by a real component (\emph{r}) and three imaginary components as follows:
% \vspace{-5pt}
\begin{equation}
\label{equa:quaternion}
X = r + a \boldsymbol{i} + b \boldsymbol{j} + c \boldsymbol{k},
\end{equation}
where $\boldsymbol{ijk} = \boldsymbol{i^2} = \boldsymbol{j^2} = \boldsymbol{k^2} = -1$. The non-commutative multiplication rules of quaternion numbers are: $\boldsymbol{ij}=\boldsymbol{k}$, $\boldsymbol{jk}=\boldsymbol{i}$, $\boldsymbol{ki}=\boldsymbol{j}$, $\boldsymbol{ji}=-\boldsymbol{k}$, $\boldsymbol{kj}=-\boldsymbol{i}$, $\boldsymbol{ik}=-\boldsymbol{j}$. In Equa~(\ref{equa:quaternion}), $r, a, b, c$ are real numbers $\in \mathbb{R}$. Note that we can extend $r, a, b, c$ to real-valued vectors to obtain a Quaternion embedding, which we use to represent each user/item's latent features and conduct neural transformations. Operations on Quaternion embeddings are similar to Quaternion numbers.

%Next, we introduce Quaternion-based operations on Quaternions numbers/embeddings as follows: 
% Note that the operations are applied for both Quaternion numbers and Quaternion embeddings.

\noindent\textbf{Component-wise Quaternion Operators:} Let $f$ define an algebraic operator in real space $\mathbb{R}$. The \emph{component-wise Quaternion operator} $f$ on two Quaternions $X, Y \in \mathbb{H}$ is defined as:
\begin{equation}
\label{equa:quaternion-component-wise}
f(X, Y) = f(r_X, r_Y) +
        f(a_X, a_Y)\boldsymbol{i} +
        f(b_X, b_Y)\boldsymbol{j} +
        f(c_X, c_Y)\boldsymbol{k}
\end{equation}
For instance, if $f$ is an \emph{addition} operator (i.e. $f(a, b) = a + b$), then $f(X, Y)$ returns a component-wise Quaternion addition between $X$ and $Y$. If $f$ is a \emph{dot product} operator (i.e. $f(a, b)=a^Tb$), then $f(X,Y)$ returns a component-wise Quaternion \emph{dot product} between $X$ and $Y$. A similar description is applied when $f$ is either \emph{subtraction}, \emph{scalar multiplication}, \emph{product}, \emph{softmax}, or \emph{concatenate} operator, \emph{.etc}.

\noindent\textbf{Hamilton Product:} The Hamilton product (denoted by the $\otimes$ symbol) of two Quaternions $X \in \mathbb{H}$ and $Y \in \mathbb{H}$ is defined as:
% \vspace{-2pt}
\begin{equation}
\label{equa:quaternion-hamiltonprod}
\begin{aligned}
X \otimes Y =& (r_X r_Y - a_X a_Y - b_X b_Y - c_X c_Y) \; + \\
             & (r_X a_Y + a_X r_Y + b_X c_Y - c_X b_Y) \boldsymbol{i} \;+ \\
             & (r_x b_Y - a_X c_Y + b_X r_Y + c_X a_Y)  \boldsymbol{j} \;+ \\
             & (r_X c_Y + a_X b_Y - b_X a_Y + c_X r_Y)  \boldsymbol{k}  \\
\end{aligned}
% \vspace{-3pt}
\end{equation}
%Intuitively, the Hamilton product represents inter-dependency interactions between two input Quaternions. When the inputs are two Quaternion embeddings, the Hamilton product will encode the latent inter-dependencies between two groups of input features. Hence, we will use the Hamilton product extensively for transforming Quaternion embeddings in our proposals. This differs to the real-space embeddings where the ordinary product is mostly used in vector and matrix transformation.

%Hamilton product is the heart of Quaternions, and is used extensively in this paper.
%For an easy reference, we provide additional operations on Quaternion that we will use in the paper as follows:

\noindent\textbf{Activation function on Quaternions:} Similar to \cite{parcollet2018quaternion,gaudet2018deep}, we use a \emph{split activation function} because of its stability and simplicity. \emph{Split activation function} $\beta$ on a Quaternion $X$ is defined as:
% \vspace{-4pt}
\begin{equation}
\label{equa:quaternion-activation}
\beta(X) = \alpha(r)  + \alpha(a) \boldsymbol{i} + \alpha(b)\boldsymbol{j} + \alpha(c)\boldsymbol{k}
\end{equation}
, where $\alpha$ is any standard activation function for real values.

\noindent\textbf{Concatenate four components of a Quaternion:} 
concatenates all four Quaternion components into one real-valued vector:
%simply concatenate the real component value and the three imaginary components' values into one real-valued vector:
% \vspace{-3pt}
\begin{equation}
\label{equa:quaternion-concate-single}
\begin{aligned}
[X] = [r_X, a_X, b_X, c_X]
\end{aligned}
\end{equation}

%% file: 4-model.tex
\section{Our proposed models}
\label{sec:models}
\begin{figure*}
    \centering
    \includegraphics[width=0.73\textwidth]{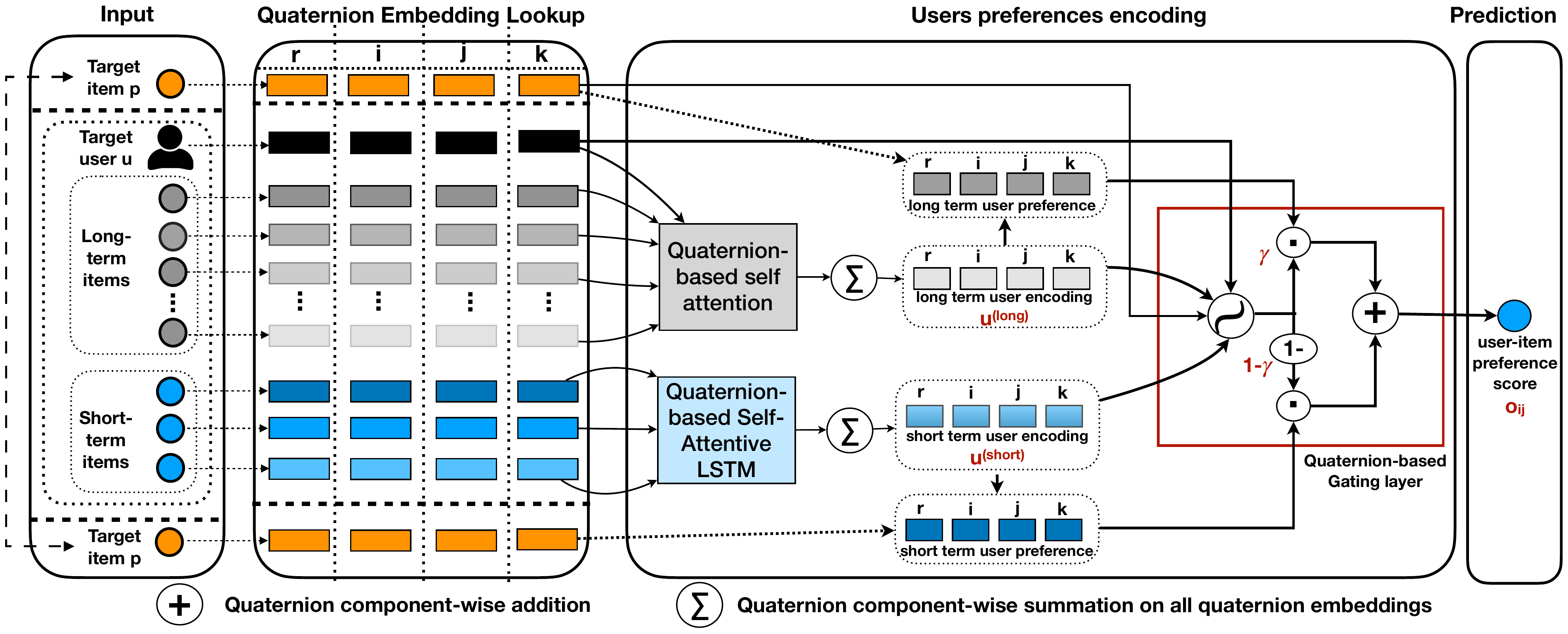}
    \vspace{-12pt}
    \caption{Our proposed architecture for modeling both long and short-term user interests using Quaternion representations.}
    \label{fig:model-architecture}
    \vspace{-12pt}
\end{figure*}

Figure \ref{fig:model-architecture} shows an overview of our proposals. First, our QUaternion-based self-Attentive Long term user Encoding (\emph{QUALE}) learns a user's long-term interest by using long-term consumed items and the target item. Second, our QUaternion-based self-Attentive Short term user Encoding (\emph{QUASE}) encodes the user's short-term intent by using short-term consumed items and the target item. Then our QUaternion-based self-Attentive Long Short term user Encoding (\emph{QUALSE}) fuses both of the user preferences by using a Quaternion-based gating layer. We describe each component as follows: %in the following subsections in detail.

% long-term consumed items are fetched into the long-term user-interest encoding with the help of the proposed Quaternion-based self-Attention model, and output a long-term user encoding. Then, the long-term user encoding is fused with the target item\'s Quaternion embedding to produce the long-term user preference. Second, the short-term consumed items are fetched into our proposed Quaternion-based self-Attentive LSTM to learn the user's short-term encoding. Next, the user's short-term encoding is combined with the target item Quaternion embedding to produce a short-term user preference. Third, we propose a Quaternion-based gating layer to combine both the user's long-term and short-term preferences. We describe each component in details as follows:

%%%%%%%%%%%%%%%%%%%%%%%%%%%%%% QUAL MODEL %%%%%%%%%%%%%%%%%%%%%%%%%%%%%%%%%%%%%%%%%%%
% \vspace{-5pt}
\subsection{QUaternion-based self-Attentive Long term user encoding (our QUALE model)}
%In this subsection, we aim to model the long-term or general user interests that are supposed to be changed slowly.
% The user's long-term interests capture the long-range relationships between the earlier consumed items and the target item. 
%Importance and effectiveness of the long-term interests have been addressed in \cite{koren2008factorization,belletti2019quantifying,he2016fast}.
The most widely used technique for modeling the user long-term interests is the Asymmetric-SVD (ASVD) \cite{koren2008factorization} model. Its basic idea is to encode each user and item by latent representations where the user representation is encoded by summing latent representations of the user's interacted items.
%The biggest advantage of ASVD over other general recommenders \cite{he2017neural,hu2008collaborative,tay2018latent,he2016fast} lays in its ability to produce latent representations for non-existed users with some observed interactions without rebuilding the model.
To an extent, we propose a QUaternion-based self-Attentive Long term user Encoding (\emph{QUALE}). \emph{QUALE} represents each user and each item as Quaternion embeddings. Then, we encode each user by attentively summing Quaternion embeddings of her interacted items as follows:
%To an extent, we propose a Quaternion-based self-attentive version of ASVD, named as QUaternion-based self-Attentive Long term user Encoding (\emph{QUALE}). \emph{QUALE} represents each user and each item as Quaternion embeddings. Then, we encode each user by attentively summing Quaternion embeddings of her interacted items as follows:
\vspace{-2pt}
\begin{equation}
\label{equa:qasvd}
\boldsymbol{u_i}^{(long)} = \sum^{|L^{(u_i)}|}_{k=1} \alpha_k \times  \boldsymbol{p_k}^{(long)}
\vspace{-2pt}
\end{equation}
where $\boldsymbol{u_i}^{(long)}, \boldsymbol{p_k}^{(long)} \in \mathbb{H}$. 
The summation ``$\sum$'' and the multiplication ``$\times$'' are Quaternion component-wise operators, which are calculated by using Equa~(\ref{equa:quaternion-component-wise}).
We use our proposed Quaternion personalized self-attention mechanism to assign attentive scores $\alpha_k \in \mathbb{H}$ for different long-term items $p_k$.

Our \emph{QUALE} model has four layers: Input, Quaternion Embedding, Encoding, and Output layers. We detail each layer as follows:

% \vspace{-2pt}
\subsubsection{Input} \emph{QUALE} requires a target user $u_i$, a target item $p_j$, and the user's list of $l$ long-term items $L^{(u_i)}$ with $|L^{(u_i)}| = l$. $l$ could be simply set to the maximum number of long-term items among all the users in a dataset. However, we observed that only several users in our datasets consumed an extremely large number of items compared to the majority of users. Hence, we set $l$ to the upper bound of the boxplot approach (i.e. Q3 + 1.5IQR, where Q3 is the \emph{third quartile}, and IQR is the \emph{Interquartile range} of the sequence length distribution of all users). If a user has consumed less than $l$ items, we pad the list with \emph{zeroes} until its length reaches $l$.

%\subsubsection{Input} \emph{QUALE} requires a target user $u_i$, a target item $p_j$, and the user's list of $l$ long-term items $L^{(u_i)}$ with $|L^{(u_i)}| = l$, where $l$ represents the maximum number of long-term items among all the users in a dataset. If a user has consumed less than $l$ items, we pad the list with \emph{zeroes} until its length reaches $l$.

% \vspace{-2pt}
\subsubsection{Quaternion Embedding layer} It holds two Quaternion embedding matrices: a user context Quaternion embedding matrix $\mathcal{U}^{(long)} \in \mathbb{H}^{m \times d}$, and an item Quaternion embedding matrix $\mathcal{P}^{(long)}$ $\in$ $\mathbb{H} ^{n \times d}$. Here, $m$ and $n$ are the respective number of users and items in the system. $d$ is the Quaternion embedding size, and is measured by the total size of real-valued vectors of four Quaternion components ($d = |r| + |a| + |b| + |c|$, and $|r| = |a| = |b| = |c| = d/4$). 
By passing the target user $u_i$, the target item $p_j$, and long-term items $p_k$ in the \emph{Input} layer through the two respective Quaternion embedding matrices, we obtain the corresponding user context Quaternion embedding $\boldsymbol{q_i}^{(long)}$,  target item Quaternion embedding $\boldsymbol{p_j}^{(long)}$ and long-term item Quaternion embeddings $\boldsymbol{p_k}^{(long)}$.
\subsubsection{Encoding layer} Its main goal is to compute attentive scores for $l$ Quaternion item embeddings in Equa~(\ref{equa:qasvd}). To do so, we propose a Quaternion personalized self-attention mechanism as follows:

We first compute the Hamilton product between each long-term item Quaternion embedding $\boldsymbol{p_k}^{(long)}$ ($k=\overline{1, l}$) and the Quaternion context embedding $\boldsymbol{q_i}^{(long)}$ of the target user $u_i$. Next, we use Equa~(\ref{equa:quaternion-component-wise}) to multiply the results with the scaling factor $1 / \sqrt{d}$ to eliminate the scaling effects. Then, we apply Component-wise Softmax (Equa~(\ref{equa:quaternion-component-wise})) to obtain Quaternion attention scores as follows:
% \vspace{-5pt}

\begin{equation}
\label{equa:q-att-scores}
\resizebox{0.36\textwidth}{!}{$
\begin{aligned}
        \begin{bmatrix}
        {\alpha_{1}} \\
        {\alpha_{2}} \\
        \dots \\
        {\alpha_{l}} \\
        \end{bmatrix}
        =
        \text{ComponentSoftmax}
        \begin{pmatrix}
        \begin{bmatrix}
        \boldsymbol{p_1}^{(long)} \otimes \boldsymbol{q_i}^{(long)} / \sqrt{d} \\
        \boldsymbol{p_2}^{(long)} \otimes \boldsymbol{q_i}^{(long)} / \sqrt{d}\\
        \dots \\
        \boldsymbol{p_l}^{(long)} \otimes \boldsymbol{q_i}^{(long)} / \sqrt{d}\\
        \end{bmatrix}
        \end{pmatrix}
        % =
        % \begin{bmatrix}
        % \boldsymbol{r_{1}} \\
        % \boldsymbol{r_{2}} \\
        % \dots \\
        % \boldsymbol{r_{l}} \\
        % \end{bmatrix}
        % +
        % \begin{bmatrix}
        % \boldsymbol{a_{1}} \\
        % \boldsymbol{a_{2}} \\
        % \dots \\
        % \boldsymbol{a_{l}} \\
        % \end{bmatrix} \boldsymbol{i}
        % +
        % \begin{bmatrix}
        % \boldsymbol{b_{1}} \\
        % \boldsymbol{b_{2}} \\
        % \dots \\
        % \boldsymbol{b_{l}} \\
        % \end{bmatrix} \boldsymbol{j}
        % +
        % \begin{bmatrix}
        % \boldsymbol{c_{1}} \\
        % \boldsymbol{c_{2}} \\
        % \dots \\
        % \boldsymbol{c_{l}} \\
        % \end{bmatrix} \boldsymbol{k}
\end{aligned}
% \vspace{-5pt}
$}
\end{equation}
% , where $e_{1}, e_{2}, ..., e_{1} \in \mathbb{H}$ are Quaternion vectors
%, $r_{1}, r_{2}$, $..., r_{l} \in \mathbb{R}$ are real-valued vectors of the Quaternion real component, and $a_{1}, a_{2}, ..., a_{l} \in \mathbb{R}$, $b_{1}, b_{2}, ..., b_{l} \in \mathbb{R}$, $c_{1}, c_{2}, ..., c_{l} \in \mathbb{R}$ are the corresponding real-valued vectors of the three Quaternion imaginary components
.% Next, we divide the results in Equa~(\ref{equa:q-att-raw}) by $\sqrt{d}$ to eliminate the scaling effects and use the Component-wise Softmax in Equa~(\ref{equa:quaternion-softmax}) to

To obtain the attentive long-term user encoding $\boldsymbol{u_i}^{(long)}$ of the user $u_i$, we first perform the component-wise product 
% (using Equa~(\ref{equa:quaternion-component-wise})) 
between the attention scores $[\alpha_{1}, \alpha_{2}, ..., \alpha_{l}]$ obtained in Equa~(\ref{equa:q-att-scores}) with its corresponding item Quaternion embeddings $[p_1^{(long)}, p_2^{(long)},$ $..., p_l^{(long)}]$. Then we sum them up to obtain $\boldsymbol{u_i}^{(long)}$ as follows:

%To obtain the attentive encoding of the user $u_i$, we first perform the component-wise product (using Equa~(\ref{equa:quaternion-prod})) between the attention scores $[\alpha_{1}, \alpha_{2}, ..., \alpha_{l}]$ obtained in Equa~(\ref{equa:q-att-scores}) with its corresponding item Quaternion embeddings $[p_1^{(long)}, p_2^{(long)}, ..., p_l^{(long)}]$, and then we sum them up to obtain an attentive Quaternion embedding $\boldsymbol{u_i}^{(long)}$ (i.e., long term user encoding as shown in Figure~\ref{fig:model-architecture}):

\vspace{-4pt}
\begin{equation}
\label{equa:q-user-long}
    \boldsymbol{u_i}^{(long)} = \sum_{k=1}^{l} \alpha_k \times \boldsymbol{p_k}^{(long)}
\vspace{-2pt}
\end{equation}

\noindent\textbf{Our proposed Quaternion personalized self-attention mechanism vs. the existing self-attention mechanism:} Our proposed Quaternion personalized self-attention mechanism is different from the self-attention mechanism that has been widely used in the NLP tasks in two aspects. First, unlike the prior work \cite{zhong2018global}, which uses a single global context to assign attentive scores for different dialogue states, our attention mechanism provides personalized contexts for different users. In the recommendation domain, the long-term/general user interests are supposed to be changed slowly, but user interests are various across users. In other words, a user's long-term context is quite static, but different from another user. Hence, using personalized contexts for different users is better than using a single global context, which is not personalized. Second, our attention mechanism adopts Hamilton product and works for Quaternion embeddings as input, instead of the real-valued embeddings like traditional self-attention mechanisms.

% \vspace{-5pt}
\subsubsection{Output} We produce a long-term preference score  $o^{(long)}_{ij}$ between the target user $u_i$ and the target item $p_j$ by computing the Component-wise dot product between the user long-term Quaternion encoding $\boldsymbol{u_i}^{(long)}$ obtained in Equa~(\ref{equa:q-user-long}) and the target item Quaternion embedding $\boldsymbol{p_j}^{(long)}$. This results in a Quaternion score
%where the value of each component is a scalar real number
. To obtain a real-valued scalar preference score used in the parameter estimation phase, we compute the average of the scalar values of four Quaternion components by following \cite{zhang2019quaternion}:
% \vspace{-1pt}

\begin{equation}
\label{equa:q-output-long}
o^{(\text{long})}_{ij} = \text{Average}(\text{ComponentDot}(\boldsymbol{u_i}^{(long)} , \boldsymbol{p_j}^{(long)}))
% \vspace{-2pt}
\end{equation}

%%%%%%%%%%%%%%%%%%%%% SHORT-TERM MODELING %%%%%%%%%%%%%

% \vspace{-2pt}
\subsection{QUaternion-based self-Attentive Short term user Encoding (our QUASE model)}
%We use models in Quaternion Recurrent Neural Network class to model the user short-term interest.
% RNN-based models have recently gained a lot of attention because of their capability to capture item-to-item relationships \cite{zhou2019deep,wu2017recurrent,okura2017embedding}. However, due to its limitation in modeling a long sequence, we only exploit the RNN architecture to encode a user's short-term interests. Recently, \cite{parcollet2018quaternion} has introduced a Quaternion LSTM model and has shown its efficiency and effectiveness over a traditional real-valued LSTM model. However, similar to the real-valued LSTM model, the proposed Quaternion LSTM used only the last hidden state as a latent summary of the input. Instead, it is better to use all hidden states with corresponding attentive scores since this approach performed better than the original LSTM approach only using the last hidden state \cite{li2017neural} in NLP and recommendation tasks. Following this idea, we propose a Quaternion-based self-Attentive LSTM model to learn a user's short-term interest. We name our proposal as a QUaternion-based self-Attentive Short term user Encoding (QUASE). \emph{QUASE} has 4 layers: Input, Quaternion Embedding, Encoding, and Output layers. We describe each layer as follows:
RNN-based models have gained a lot of attention because of their capability to capture item-to-item relationships \cite{zhou2019deep,wu2017recurrent,okura2017embedding}. However, due to its limitation in modeling a long sequence, we only exploit the RNN architecture to encode a user's short-term interest. Recently, \cite{parcollet2018quaternion} has introduced a Quaternion LSTM (QLSTM) model and has shown its efficiency and effectiveness over a traditional real-valued LSTM model. 
However, QLSTM used only the last hidden state as a latent summary of the input, which is suboptimal. To an extent, we propose a Quaternion-based self-Attentive LSTM model to learn a user's short-term interest. We name our proposal as a QUaternion-based self-Attentive Short term user Encoding (QUASE). \emph{QUASE} has 4 layers: Input, Quaternion Embedding, Encoding, and Output layers. We describe each layer as follows:

% However, similar to the real-valued LSTM model, the proposed Quaternion LSTM used only the last hidden state as a latent summary of the input. Instead, it is better to use all hidden states with corresponding attentive scores since this approach performed better than the original LSTM approach only using the last hidden state \cite{li2017neural} in NLP and recommendation tasks. Following this idea, we propose a Quaternion-based self-Attentive LSTM model to learn a user's short-term interest. We name our proposal as a QUaternion-based self-Attentive Short term user Encoding (QUASE). \emph{QUASE} has 4 layers: Input, Quaternion Embedding, Encoding, and Output layers. We describe each layer as follows:

%Among all RNN based models (i.e. LSTM, GRU), LSTM and GRU models are the most popular ones. We observe that LSTM model works better than GRU, so our QUASE model use it.
%Recently, \cite{parcollet2018quaternion} has introduced a Quaternion LSTM model

% \vspace{-5pt}
\subsubsection{Input} A target item $p_j$, and the chronological list of $s$ short-term consumed items $S^{(u_i)}$ of the target user $u_i$ with $|S^{(u_i)}| = s$, where $s$ represents the maximum number of short-term items among all the users in a dataset. If a user has consumed less than $s$ items, we pad the list with \emph{zeroes} until its length reaches $s$.

% \vspace{-5pt}
\subsubsection{Quaternion Embedding layer} It holds an item Quaternion Embedding matrix $\mathcal{P}^{(short)}$ $\in \mathbb{H}^{n \times d}$. By passing the target item $p_j$, and $s$ short-term items in the $S^{(u_i)}$ of the target user $u_i$ through $P^{short}$, we obtain their corresponding Quaternion embeddings $\boldsymbol{p_j}^{(short)}$, and \{$\boldsymbol{p_1}^{(short)}, \boldsymbol{p_2}^{(short)}, ..., \boldsymbol{p_s}^{(short)} $\}.

% \vspace{-5pt}
\subsubsection{Encoding layer}
%The main goal of this layer is to encode the user's short-term interests by discovering the sequential transition patterns from short-term consumed items to the target item. \cite{parcollet2018quaternion} has recently shown better effectiveness of a Quaternion-based LSTM model over the traditional real-valued LSTM model in NLP applications. Therefore, we adapt the Quaternion-based LSTM into modeling the item-item sequential transition as follows:
In this layer, we adapt the recently introduced Quaternion-based LSTM to model the item-item sequential transition.
Denote $\boldsymbol{p_t^{(short)}}$ is the Quaternion embedding of the $t^{th}$ short-term item $p_t \in S^{(u_i)}$ ($t=\overline{1, s}$). Let $f_t, i_t, o_t, c_t$, and $h_t$ be the forget gate, input gate, output gate, cell state, and the hidden state of a Quaternion LSTM cell at time step $t$, respectively. We compute these variables as follows:
% \vspace{-3pt}
\begin{equation}
\label{equa:qlstm}
\begin{aligned}
f_t &= \sigma (W_f \otimes \boldsymbol{p_t}^{(short)} + R_f \otimes  \boldsymbol{h_{t-1}} + g_f)  \\
i_t &= \sigma (W_i \otimes \boldsymbol{p_t}^{(short)} + R_i \otimes  \boldsymbol{h_{t-1}} + g_i ) \\
o_t &= \sigma (W_o \otimes \boldsymbol{p_t}^{(short)} + R_o \otimes  \boldsymbol{h_{t-1}} + g_o ) \\
c_t &= f_t \times c_{t-1} + i_t \times \text{tanh}(W_c \otimes \boldsymbol{p_t}^{(short)} + R_c \otimes  \boldsymbol{h_{t-1}} + g_c ) \\
h_t &= o_t \times tanh(c_t)
\end{aligned}
\end{equation}
, where $W_f, R_f, W_i, R_i, W_o, R_o, W_c, R_c$ are Quaternion weight matrices. $g_f, g_i, g_o, g_c$ are Quaternion bias vectors. $f_t, i_t, o_t, c_t, h_t$ are Quaternion vectors. 
The ``$\times$'' sign denotes a component-wise \emph{product} operator, which is calculated using Equa~(\ref{equa:quaternion-component-wise}).  \emph{sigmoid} $\sigma$ and \emph{tanh} are split activation functions and are computed using the Equa~(\ref{equa:quaternion-activation}).

Using Equa~(\ref{equa:qlstm}), given $s$ short-term consumed items $p_1, p_2, ..., p_s$, we obtain their respective output Quaternion hidden states $\boldsymbol{h_1}, \boldsymbol{h_2},$ $..., \boldsymbol{h_s}$.
Then, we propose a Quaternion self-attention mechanism to combine all $s$ output Quaternion hidden states before using it to predict the next item.
%To filter out the effects of irrelevant items in real-valued RNN models, previous works have adopted real-valued attention mechanisms to suppress the information that deviated from the target prediction, while encouraging relevant items with higher influence scores \cite{wang2018attention,zhou2019deep,ying2018sequential,yu2019adaptive}. Therefore, rather than using the last Quaternion hidden state as the user's short-term representation, we propose a Quaternion self-attention mechanism to combine all output Quaternion hidden states before using it to predict the next item.
Different from the long-term user preferences where they are supposed to be static or changed very slowly, the short-term user interests are dynamic and changed quickly. Hence, using a static user context for each user to make personalized attention like what we did for the \emph{QUALE} model is not ideal. Instead, we define a Quaternion global context vector to capture the sequential transition patterns from item to item among all the users. Denote $q$ as a Quaternion global context vector, the Quaternion-based self-attention score of each hidden state $h_t$ is measured by:
\vspace{-2pt}
\begin{equation}
\label{equa:q-short-att-scores}
\begin{aligned}
\begin{bmatrix}
\alpha_1^{(short)}  \\
\alpha_2^{(short)} \\
\dots \\
\alpha_s^{(short)} \\
\end{bmatrix}
= \text{ComponentSoftmax}
\begin{pmatrix}
\begin{bmatrix}
{\boldsymbol{h_1} \otimes \boldsymbol{q}} / {\sqrt{d}}  \\
{\boldsymbol{h_2} \otimes \boldsymbol{q}} / {\sqrt{d}}  \\
\dots \\
{\boldsymbol{h_t} \otimes \boldsymbol{q}} / {\sqrt{d}}  \\
\end{bmatrix}
\end{pmatrix}
\end{aligned}
\vspace{-3pt}
\end{equation}
, where $\alpha_1^{(short)}, \alpha_2^{(short)}, ..., \alpha_s^{(short)}$ are Quaternion numbers. To achieve the final short-term user Quaternion encoding, we perform a component-wise product 
% (using Equa~(\ref{equa:quaternion-prod})) 
between the Quaternion hidden states and their respective Quaternion attention scores, followed by a Hamilton product with a Quaternion weight matrix $W$ and the split activation function \emph{tanh}:
\vspace{-3pt}
\begin{equation}
\label{equa:q-user-short}
\boldsymbol{u_i}^{(short)} = tanh \bigg( W \otimes \bigg(
        \sum_{t=1}^{s} \alpha_t^{(short)} \times \boldsymbol{h_t}
\bigg) \bigg)
\vspace{-2pt}
\end{equation}

%Note that previous works shown that using GRU cell \cite{cho2014learning} in recurrent neural networks could gain competitive performance compared to using LSTM cell. Thus, we also further extend our proposed Quaternion self-Attentive LSTM to a Quaternion self-Attentive GRU by replacing the Equa~(\ref{equa:qlstm}) with:

Note that we also designed a Quaternion self-Attentive GRU 
% using GRU cell \cite{cho2014learning}
% by using GRU cell \cite{cho2014learning}
, but its performance was slightly worse than the Quaternion self-Attentive LSTM (see Table \ref{table:PerformanceComparison} in Section \ref{sec:exp}). Thus, we only described the Quaternion self-Attentive LSTM due to space limitation.
% Note that using GRU cell \cite{cho2014learning} in recurrent neural networks is also popular. So, we further  extend our proposed Quaternion self-Attentive LSTM to a Quaternion self-Attentive GRU by replacing the Equa~(\ref{equa:qlstm}) with:
% \vspace{-\left( }
% \begin{equation}
% \label{equa:qgru}
% \begin{aligned}
% z_t =& \sigma (W_z \otimes \boldsymbol{p_t}^{(short)} + R_z \otimes  \boldsymbol{h_{t-1}} + g_z)  \\
% f_t =& \sigma (W_f \otimes \boldsymbol{p_t}^{(short)} + R_f \otimes  \boldsymbol{h_{t-1}} + g_f ) \\
% h_t =& z_t \times h_{t-1} \; \; + \\
%      &(1-z_t) \times tanh \big( W_h \otimes \boldsymbol{p_t}^{(short)} + R_h \otimes (f_t \times \boldsymbol{h_{t-1}}) + g_h \big)
% \end{aligned}
% \end{equation}
% , where $z_t, f_t$ are update gate and reset gate, respectively. ``$\times$'' denotes a Quaternion component-wise product. $W_z, R_z, W_f, R_f, W_h, R_h$ are Quaternion weight matrices. $g_z, g_f, g_h$ are Quaternion bias vectors. The user's long-term encoding using the Quaternion GRU and the Quaternion self-attention mechanism can be obtained in the same manner by using Equa~(\ref{equa:q-short-att-scores}) and (\ref{equa:q-user-short}).

% \vspace{-5pt}
\subsubsection{Output}
Similar to Equa~(\ref{equa:q-output-long}), we produce the user $u_i$ short-term preference score $o_{ij}^{(short)}$ over the target item $p_j$ as follows:
%by compute the component-wise dot product between the user's short-term Quaternion encoding obtained in Equa~(\ref{equa:q-user-short}) with the target item Quaternion embedding $\boldsymbol{p_j}^{(short)}$, then averaging values of four components in the resulted Quaternion score:
%To produce a short-term preference score $o_{ij}^{(short)}$ between the target user $u_i$ and the target item $p_j$, we compute the component-wise dot product (using Equa~(\ref{equa:quaternion-dotprod})) between the user's short-term Quaternion encoding obtained in Equa~(\ref{equa:q-user-short}) with the target item Quaternion embedding $\boldsymbol{p_j}^{(short)}$. This, again, leads to a Quaternion score where values of its four components are four scalar real numbers. To achieve a scalar real-valued score, we average the four component's values of the resulted Quaternion score:
\vspace{-5pt}
\begin{equation}
\label{equa:q-output-short}
o_{ij}^{(short)} = Average \big(
        ComponentDot ( \boldsymbol{u_i}^{(short)}, \boldsymbol{p_j}^{(short)})
        \big)
\end{equation}

%%%%%%%%%%%%%%%%%%%% FUSION OF TWO MODELS. %%%%%%%%%%%%%%%%%%%%%%%%%%%%%%
% \vspace{-7pt}
\subsection{QUaternion-based self-Attentive Long Short term user Encoding (QUALSE): a Fusion of QUASE and QUALE models}
In this part, we aim to combine both user's long-term and short-term preferences modeling parts into one model, namely \emph{QUALSE}, fusing \emph{QUALE} and \emph{QUASE} models. 
% In this part, we aim to combine both user's long-term and short-term preferences modeling parts into one model, namely \emph{QUALSE}. In the previous section, we showed that the short-term user encoding part can be modeled by using either Quaternion LSTM or Quaternion GRU. However, in our experiments presented in the following section, we observe that using Quaternion LSTM gains slightly better performance compared to Quaternion GRU. Hence, we opt to Quaternion LSTM in \emph{QUASE} model when fusing \emph{QUALE} and \emph{QUASE} models.
Inspired by the gated mechanism in LSTM \cite{hochreiter1997long} to balance the contribution of the current input and the previous hidden state, we propose a \emph{personalized} Quaternion gated mechanism to fuse the long-term and short-term user interests learned in \emph{QUALE} and \emph{QUASE} models. Our \emph{personalized} gating proposal is different to the traditional gating mechanism in two folds. First, gating weights in our proposal are in Quaternion space and the transformations are computed using the Hamilton product.
Second, as users' behaviors differ from a user to another user, we additionally input the target user embeddings $\boldsymbol{u_i}$ to let the gating layer assign personalized scores for different users.
The long-term and short-term interest fusion is computed as follows:
%Specifically, the interest fusion between the target user $u_i$ and the target item $p_j$, \emph{w.r.t} both long-term and short-term preference aspects, is computed as follows:
% \vspace{-2pt}
\begin{equation}
\label{equa:q-output-combine}
\resizebox{0.43\textwidth}{!}{$
\begin{aligned}
    \gamma_{ij}^{(long)} =& \sigma
                            \big(
                                 W_g^{(1)} \otimes [\boldsymbol{u_i}^{(long)}, \boldsymbol{u_i}^{(short)}] +
                                W_g^{(2)} \otimes \boldsymbol{u_i} +
                                W_g^{(3)} \otimes \boldsymbol{p_j}
                            \big) \\
    o_{ij} =& W_o^{(1)}
            \big[\gamma_{ij}^{(long)} \times  (\boldsymbol{u_i}^{(long)} \times \boldsymbol{p_j}^{(long)}) \big] \; + \\
            & W_o^{(2)}
            \big[(1-\gamma_{ij}^{(long)}) \times (\boldsymbol{u_i}^{(short)} \times \boldsymbol{p_j}^{(short)}) \big]
\end{aligned}
$}
% \vspace{-3pt}
\end{equation}
, where $W_g^{(1)},  W_g^{(2)}$, and $W_g^{(3)}$ are Quaternion weight matrices, $\boldsymbol{u_i}^{(long)}$ and $\boldsymbol{u_i}^{(short)}$ are the user's long-term Quaternion encoding and short-term Quaternion encoding obtained in Equa~(\ref{equa:q-user-long}) and (\ref{equa:q-user-short}), respectively. $[\cdot \;, \cdot]$ is the component-wise concatenate (Equa~(\ref{equa:quaternion-component-wise})) of two input Quaternion vectors. To compute the long-term gate $\gamma_{ij}^{(long)}$, $\boldsymbol{u_i}$ and $\boldsymbol{p_j}$ are introduced as an additional user context Quaternion embedding and a target item context Quaternion embedding to let the model know which long-term or short-term interests are more relevant. To measure the final output $o_{ij}$, since $\gamma_{ij}^{(long)}$ is a Quaternion vector while $o_{ij}^{(long)}$ and $o_{ij}^{(short)}$ are scalar values, we reconstruct the user's long-term interest by computing $\boldsymbol{u_i}^{(long)} \times \boldsymbol{p_j}^{(long)}$ and the short-term interest by measuring $\boldsymbol{u_i}^{(short)} \times \boldsymbol{p_j}^{(short)}$, which are also Quaternion vectors. Finally, to combine multiple dimensional features from the weighted long-term and short-term interest Quaternion vectors, we concatenate all their components, denoted by $[\cdot]$ (Equa~(\ref{equa:quaternion-concate-single})), and use two real-valued weight vectors $W_o^{(1)}$ and $W_o^{(2)}$ to produce a fused preference score as a scalar real number. Note that in \emph{QUALSE}, \emph{QUASE} and \emph{QUALE} hold separated item memory to increase the their flexibility.

%\begin{figure}
%    \centering
%    \includegraphics[width=\linewidth]{figs/QuaternionAdversarial.pdf}
%    \vspace{-15pt}
%    \caption{Our Quaternion Adversarial attack on BPR. }
%    \label{fig:q-adv-loss-figure}
%    \vspace{-15pt}
%\end{figure}

% \vspace{-5pt}
\subsection{Parameter Estimation}
\subsubsection{Training with Bayesian Personalized Ranking (BPR) loss} 
Given a Quaternion matrix $E \in \mathbb{H}^{(m+n)\times d}$ as the Quaternion embeddings of all users and items in the system, and $\Theta$ as other parameters of the model, 
we aim to minimize the following BPR loss function:
% training our models with BPR loss aims to minimize the following objective function:
% we adopt the widely-used BPR loss as the objective function to train our proposed models as follows:
\vspace{-3pt}
\begin{equation}
\label{equa:bpr-loss}
\resizebox{0.42\textwidth}{!}{$
\begin{aligned}
    & \mathcal{L}_{BPR} (\mathcal{D} | E, \Theta) \\
    &    = \operatorname*{\bf{argmin}}_{E, \Theta}
        \bigg (
                   - \sum_{(i, j^+, j^-)} log \sigma( o_{ij^+} - o_{ij^-}) +
                   \lambda_\Theta \Vert \Theta  \Vert _2  +
                   \lambda_E \Vert  E \Vert _2
        \bigg )
\end{aligned}
$}
\vspace{-2pt}
\end{equation}
, where ($i$, $j^{+}$, $j^{-}$) is a triplet of a target user, a target item, and a negative item that is randomly sampled from the items set $P$. $\mathcal{D}$ denotes all the training instances. $o_{ij^{+}}$ and $o_{ij^{-}}$ are the respective positive and negative preference scores, that are computed by Equa~(\ref{equa:q-output-long}), (\ref{equa:q-output-short}), (\ref{equa:q-output-combine}), corresponding to \emph{QUALE}, \emph{QUASE} and \emph{QUALSE} models. $\lambda_\Theta$ and $\lambda_E$ are regularization hyper-parameters.

\vspace{-3pt}
\subsubsection{Training with Quaternion Adversarial attacks} 
% As shown in prior works, neural networks are vulnerable to adversarial noise \cite{he2018adversarial,kurakin2016adversarial,tramer2017ensemble}. 
Previous works have shown that neural networks are vulnerable to adversarial noise \cite{he2018adversarial,kurakin2016adversarial}. 
Therefore, to increase the robustness of our models, we propose a Quaternion Adversarial attack on BPR loss, namely \emph{QABPR}. \emph{QABPR} inherits from traditional adversarial attacks for computer visions \cite{kurakin2016adversarial} and recommendation systems \cite{he2018adversarial} but differs from them: \emph{QABPR} applies for Quaternion space, while the formers apply for real-valued space. To our best of knowledge, ours is the first work using adversarial training on Quaternion space in the recommendation domain.

%Figure \ref{fig:q-adv-loss-figure} shows the general idea of our proposed \emph{QABPR}. Intuitively, during the adversarial training phase, we define learnable Quaternion perturbation noise on user and item Quaternion embeddings, and perform the Quaternion component-wise addition (Equa~(\ref{equa:quaternion-add})) to obtain crafted Quaternion embeddings for the model. The learnable Quaternion noise is optimized such that the model mis-ranks between positive items and negative items (i.e. negative items have higher preference scores than positive items). Particularly, we learn the Quaternion adversarial noise by maximizing the following cost function under the $L_2$ attack:
In \emph{QABPR}, we first define learnable Quaternion perturbation noise $\delta$ on user and item Quaternion embeddings. Then, we perform the Quaternion component-wise addition (Equa~(\ref{equa:quaternion-component-wise})) to obtain crafted Quaternion embeddings. The learnable Quaternion noise $\delta$ is optimized such that the model mis-ranks between positive items and negative items (i.e. negative items have higher preference scores than positive items). Particularly, a \emph{max} player learns $\delta$ by maximizing the following cost function under the $L_2$ attack:
% \vspace{-5pt}
\begin{equation}
\label{equa:quaternion-adv-loss}
\resizebox{0.38\textwidth}{!}{$ 	
	\begin{aligned}												
	&\mathcal{L}_{adv} (\mathcal{D} | {E}^* + \delta, \Theta^{*}) \\ 
	&= \operatorname*{\bf{argmax}}_{\delta, \Vert \delta \Vert_2 \le \epsilon}
	\bigg (
	- \sum_{(i, j^+, j^-)} log \sigma( o_{ij^+} - o_{ij^-}) + \lambda_\delta \Vert \delta \Vert _2
	\bigg )
	\end{aligned}
	$}
\vspace{-2pt}
\end{equation}
where $\epsilon$ is a noise magnitude hyper-parameter. $E^*$ and  $\Theta^{*}$ are optimal values of
%users and items Quaternion embeddings
$E$ and
%other parameters
$\Theta$ that are pre-learned in Equa~(\ref{equa:bpr-loss}) and are fixed in Equa~(\ref{equa:quaternion-adv-loss}). ${E^*} + \delta$ is the crafted Quaternion embeddings.  %Regularization parts 
$\lambda_\Theta \Vert \Theta  \Vert _2$ and $\lambda_E \Vert E \Vert_2$ in Equa~(\ref{equa:bpr-loss}) are ignored in Equa~(\ref{equa:quaternion-adv-loss}) as they become constant terms. $\lambda_\delta \Vert \delta \Vert _2$ is the noise regularization term.

Solving Equa~(\ref{equa:quaternion-adv-loss}) is expensive. 
%, especially for complicated deep neural networks
Hence, we adopt the Fast Gradient Method \cite{kurakin2016adversarial} to approximate $\delta$ as follows:
% \vspace{-3pt}
\begin{equation}
\label{equa:approx-noise}
\resizebox{0.25\textwidth}{!}{$
	\delta= \epsilon \frac{ \mathbf{\bigtriangledown}_\delta \mathcal{L}_{adv}(\mathcal{D} | E^* + \delta, \Theta^*)}
	{\Vert
		\mathbf{\bigtriangledown}_\delta \mathcal{L}_{adv}(\mathcal{D} | E^* + \delta, \Theta^*)
		\Vert_2}
	$}              
\vspace{-2pt}
\end{equation}

Then, a \emph{min} player aims to minimize the following cost functions that incorporate both non-adversarial and adversarial examples:
%Then, learning the \emph{QABPR} loss aims to minimize the following cost functions:
% \vspace{-3pt}
\begin{equation}
\label{equa:qabpr-loss}
\begin{aligned}
\mathcal{L}& _{QABPR} (\mathcal{D} | E, E + \delta^*, \Theta)  \\
&= \operatorname*{\bf{argmin}}_{E, \Theta}
\bigg(
\mathcal{L}_{BPR} (\mathcal{D} | E, \Theta) +
\lambda_{adv} \mathcal{L}_{BPR} (\mathcal{D} | E + \delta^*, \Theta)
\bigg)
\end{aligned}
\vspace{-2pt}
\end{equation}
where $\delta^*$ is the adversarial noise that is already learned in Equa~(\ref{equa:approx-noise}), and is fixed in Equa~(\ref{equa:qabpr-loss}). $\lambda_{adv}$ is a hyper-parameter to balance the effect of the partial adversarial loss. 
Training \emph{QABPR} now becomes playing a \emph{minimax} game, where the \emph{min} and \emph{max} players play alternatively. We stop the game after a fixed number of epochs (i.e. 30 epochs) and report results based on the best \emph{validation} performance. 

Note that we name our \emph{QUALE}, \emph{QUASE}, and \emph{QUALSE} trained with \emph{QABPR} loss as \emph{AQUALE}, \emph{AQUASE}, and \emph{AQUALSE} with ``A'' denotes “adversarial”, respectively.

%It is worth to note that training with \emph{QABPR} loss only increases the training time since during the inference phase, the adversarial noise $\delta$ is ignored.

%\textbf{Autograd for back propagation with our proposals} 

%% file: 5-experiments.tex
%  \vspace{-5pt}
\section{Empirical Study}
\label{sec:exp}

In this section, we design experiments to answer the following research questions:
\squishlist
\item \textbf{RQ1:} How do our proposals work compared to the baselines?
\item \textbf{RQ2:} How do a user's long-term, short-term preference encoding models and the fused model perform?
\item \textbf{RQ3:} Is using Quaternion representation helpful and why?
\item \textbf{RQ4:} Are the gating fusion mechanism and the Quaternion BPR adversarial training helpful?
\squishend

%%%%%%%%%%%%%%%% DATASET STATISTICS %%%%%%%%%%%%%%%%%%%%
\begin{table}[t]
    \setlength\tabcolsep{1 pt}
    \centering
    \tiny
    \caption{Datasets' statistics with \# of long-term items $l$.}
    \vspace{-10pt}
    \label{table:datasets}
    \resizebox{\linewidth}{!}{
        \begin{tabular}{lcccr}
            \toprule
            Dataset                      & \# of users & \# of items & \# of actions (density \%) &   \emph{l}  \\
            \midrule
            % Apps for Android            & 79,502    & 18,416    & 555,036 (0.038\%)     & 478\\
            % Cellphone Accessories       & 47,347    & 45,454    & 262,194 (0.012\%)     & 109 \\
            % % Clothing Shoes Jewelry      & 98,185    & 138,448   & 582,271 (0.004\%)     & 205 \\
            % % Digital Music               & 7,666     & 10,799    & 75,765 (0.092\%)      & \\
            % Health Care        & 53,509    & 55,759    & 363,723 (0.012\%)     & 243\\
            % Movies TV                   & 107,734   & 73,355    & 1,262,041 (0.016\%)   & 2,442\\
            % Pet Supplies                & 25,283    & 23,762    & 160,892 (0.027\%)     & 176\\
            % Sports Outdoors             & 55,477    & 67,268    & 360,501 (0.027\%)     & 328\\
            % Tools Home                  & 30,749    & 41,828    & 194,385 (0.015\%)     & 176\\
            % Toys Games                  & 36,079    & 55,641    & 251,693 (0.013\%)     & 1,112\\
            % Video Games                 & 24,429    & 20,209    & 196,661 (0.04\%)      & 856\\
            % Yelp                        & 22,015    & 21,120    & 481,284 (0.104\%)     & 930\\
            Toys Games                  & 36k    & 55k    & 251k (0.013\%)     & 1,112\\

            Cellphone Accessories       & 47k    & 45k    & 262k (0.012\%)     & 109 \\
            % Clothing Shoes Jewelry      & 98,185    & 138,448   & 582,271 (0.004\%)     & 205 \\
            % Digital Music               & 7,666     & 10,799    & 75,765 (0.092\%)      & \\
            % Health Care                 & 53k    & 55k    & 363k (0.012\%)     & 243\\
            % Movies TV                   & 107k   & 73k    & 1,262k (0.016\%)   & 2,442\\
            Pet Supplies                & 25k    & 23k    & 160k (0.027\%)     & 176\\
            % Sports Outdoors             & 55k    & 67k    & 360k (0.027\%)     & 328\\
            % Tools Home                  & 30k    & 41k    & 194k (0.015\%)     & 176\\
            Video Games                 & 24k    & 20k    & 196k (0.040\%)      & 856\\
            Apps for Android            & 79k    & 18k    & 555k (0.038\%)     & 478\\
            Yelp                        & 22k    & 21k    & 481k (0.104\%)     & 930\\

            \bottomrule
        \end{tabular}
    }
    \vspace{-16pt}
\end{table}
\begin{table*}[t]
\centering
\large
\caption{HIT@100 and NDCG@100 of all models. Best performances are in \emph{bold}, best baseline's results are \underline{underlined}. The last two lines show the relative improvement of QUALSE and AQUALSE compared to the best baseline's results.}
\vspace{-10pt}
\label{table:PerformanceComparison}
\resizebox{0.98\textwidth}{!}{
    \begin{tabular}{p{0.5pt}p{2pt} cccccccccccc}
    \toprule
& \multicolumn{1}{c}{\multirow{2}{*}{Methods}}
& \multicolumn{2}{c}{\textbf{Toys Games}}
& \multicolumn{2}{c}{\textbf{Cellphone Acc.}}
% & \multicolumn{2}{c}{\textbf{Clothing Shoes Jewelry}}
% & \multicolumn{2}{c}{\textbf{Digital Music}}
% & \multicolumn{2}{c}{\textbf{Health Care}}
% & \multicolumn{2}{c}{\textbf{Movies TV}}
& \multicolumn{2}{c}{\textbf{Pet Supplies}}
% & \multicolumn{2}{c}{\textbf{Sports Outdoors}}
% & \multicolumn{2}{c}{\textbf{Tools Home Improvement}}
& \multicolumn{2}{c}{\textbf{Video Games}}
& \multicolumn{2}{c}{\textbf{Apps for Android}}
& \multicolumn{2}{c}{\textbf{Yelp}}
%& \multirow{2}{*}{\bf{RI}}
\\
        \cmidrule{3-14} &
            & HIT & NDCG
            & HIT & NDCG
            & HIT & NDCG
            & HIT & NDCG
            & HIT & NDCG
            & HIT & NDCG
\\
    \midrule
%%%%%%%%%% LONG TERM ENCODING MODEL
\multicolumn{1}{l@{\hskip0.5pt}|}{(a)}
            &     AASVD      &
            0.4343 & 0.1809 & 0.5640 & 0.2443 & 0.5523 & 0.2307
            & 0.5503 & 0.2229 & 0.7149 & 0.3182
            & 0.7212 & 0.3580\\
\multicolumn{1}{l@{\hskip0.5pt}|}{(b)}
            &     QCF        &
            0.3869 & 0.1560 & 0.5514 & 0.2328 & 0.5319 & 0.2194
            & 0.5217 & 0.1956 & 0.6638 & 0.2864
            & 0.6774 & 0.3119
            \\
\multicolumn{1}{l@{\hskip0.5pt}|}{(c)}
            &     NeuMF++    &
            0.3969 & 0.1553 & 0.5467 & 0.2291 & 0.5255 & 0.2174
            & 0.4944 & 0.1934 & 0.6635 & 0.2791
            & 0.6810 & 0.3208\\

\multicolumn{1}{l@{\hskip0.5pt}|}{(d)}
            &     NAIS       &
            0.4331 & 0.1796 & 0.5648 & 0.2427 & \underline{0.5569} & 0.2302
            & \underline{0.5587} & 0.2303 & 0.7076 & 0.3138 & \underline{0.7277} & 0.3573 \\

%%%%%%%%% SHORT TERM ENCODING MODELS
\multicolumn{1}{l@{\hskip0.5pt}|}{(e)}
            &     FPMC       &
            0.3370 & 0.1335 & 0.4805 & 0.1970 & 0.4405 & 0.1812
            & 0.5065 & 0.1980 & 0.6659 & 0.2847
            & 0.6704 & 0.3204\\
\multicolumn{1}{l@{\hskip0.5pt}|}{(f)}
            &     AGRU       &
            0.3747 & 0.1400 & 0.5211 & 0.2030 & 0.4690 & 0.1798
            & 0.5337 & 0.1958 & 0.6969 & 0.2960
            & 0.4722 & 0.1995\\
\multicolumn{1}{l@{\hskip0.5pt}|}{(g)}
            &     ALSTM      &
            0.3886 & 0.1419 & 0.5159 & 0.2052 & 0.4630 & 0.1685
            & 0.5156 & 0.1928 & 0.7043 & 0.2883 &
            0.5644 & 0.2519 \\
\multicolumn{1}{l@{\hskip0.5pt}|}{(h)}
            &     Caser      &
            0.3889 & 0.1507 & \underline{0.5747} & 0.2289
            & 0.4786 & 0.1859
            &  0.5502 & 0.1967 & 0.7098 & 0.3124 &
            0.6718 & 0.3201\\

%%%%%%%% COMBINATION OF LONG AND SHORT TERM
% \multicolumn{1}{l@{\hskip0.5pt}|}{(i)}
%             &     DIEN       &
%             & & & & & & & & & \\

\multicolumn{1}{l@{\hskip0.5pt}|}{(i)}
            &     SASRec     &
            0.4009 & 0.1545 & 0.5579 & 0.2239 & 0.5238 & 0.2124
            & 0.5472 & 0.2107 & 0.6706 & 0.2781 & 0.7193 & 0.3381\\

\multicolumn{1}{l@{\hskip0.5pt}|}{(j)}
            &     \text{SLi-Rec} &
            0.4267 & 0.1823 & 0.5661 & 0.2387 & 0.5502 & 0.2311
            & 0.5438 & 0.2276 & 0.7062 & 0.3117 &  0.7201 & 0.3516\\

\multicolumn{1}{l@{\hskip0.5pt}|}{(k)}
            &     ALSTM+AASVD &
            \underline{0.4394} & \underline{0.1864} & 0.5701 & \underline{0.2475} & 0.5542 & \underline{0.2326}
            & 0.5502 & \underline{0.2328} & \underline{0.7173} & \underline{0.3207} &  {0.7222} & \underline{0.3594}\\

            \midrule
\multicolumn{2}{l}{\textbf{Our proposals}}
            & & & & & & & & & & & \\

            &     QUALE       &
            0.4696 & 0.1997 & 0.6042 & 0.2685 & 0.5826 & 0.2483
            & 0.5981 & 0.2503 & 0.7281 & 0.3248 & 0.7391 & 0.3723\\
            &     QUASE\;(GRU)         &
            0.4080 & 0.1632 & 0.5612 & 0.5807 & 0.2438 & 0.5413 & 0.2246
            & 0.2207 & 0.7198 & 0.3223 & 0.6917 & 0.3324\\
            &     QUASE\;(LSTM)       &
            0.4095 & 0.1664 & 0.5844 & 0.2475 & 0.5453 & 0.2263
            & 0.5591 & 0.2261 & 0.7300 & 0.3300 & 0.6929 & 0.3311\\
            &     QUALSE        &
            \bf{0.4760} & \bf{0.2043} & \bf{0.6127} & \bf{0.2777} & \bf{0.5913} & \bf{0.2539}
            & \bf{0.6018} & \bf{0.2551} & \bf{0.7373} & \bf{0.3364} & \bf{0.7442} & \bf{0.3781}\\

            \cmidrule{2-14}
            &     AQUALE       &
            0.4831 & 0.2055 & 0.6105 & 0.2748 & 0.5902 & 0.2553
            & 0.6045 & 0.2593 & 0.7346 & 0.3306 & 0.7440 & 0.3786\\

            &     AQUASE\;(LSTM)      &
            0.4495 & 0.1847 & 0.6056 & 0.2572 & 0.5520 & 0.2329
            & 0.5762 & 0.2351 & 0.7285 & 0.3292 & 0.7048 & 0.3450 \\

            &     AQUALSE       &
            \bf{0.4921}	& \bf{0.2098}  & \bf{0.6204} & \bf{0.2842} & \bf{0.6011}  & \bf{0.2612} & \bf{0.6137} & \bf{0.2605} & \bf{0.7477} & \bf{0.3440} & \bf{0.7448} & \bf{0.3814} \\

            \cmidrule{2-14}

\multicolumn{2}{l}{\textbf{Imprv. of QUALSE}} &
            \bf{+8.33\%} & \bf{+9.60\%} &
            \bf{+6.61\%} & \bf{+12.20\%} &
            \bf{+6.18\%} & \bf{+9.16\%} &
            \bf{+7.71\%} & \bf{+9.58\%} &
            \bf{+2.79\%} & \bf{+4.90\%} &
            \bf{+2.27\%} & \bf{+5.20\%}\\
\multicolumn{2}{l}{\textbf{Imprv. of AQUALSE }}
            & \bf{+11.99\%} & \bf{+12.55\%}
            & \bf{+7.95\%} & \bf{+14.83\%}
            & \bf{+7.94\%} & \bf{+12.30\%}
            &  \bf{+9.84\%} & \bf{+11.90\%}
            & \bf{+4.24\%} & \bf{+7.27\%}
            & \bf{+2.35\%} & \bf{+6.12\%} \\
            % & & & & & & & & & \\
\bottomrule
\end{tabular}
}
\vspace{-10pt}
\end{table*}
%%%%%%%%%%%%%% END OVERALL PERFORMANCE TABLE

\vspace{-8pt}
\subsection{Datasets}
We evaluate all models on \textbf{six} public benchmark datasets collected from two real world systems as follows:
%\begin{itemize}
\squishlist
    %\item {\bf Amazon} \cite{he2016ups}: A series of datasets crawled from \emph{Amazon.com}. We use 9 different categories: \emph{Apps for Android}, \emph{Cellphone Accessories}, \emph{Health \& Personal Care}, \emph{Movies TV}, \emph{Pet Supplies}, \emph{Sports \& Outdoors}, \emph{Tools and Home Improvement}, \emph{Toys and Games}, and \emph{Video Games}.
    % \item {\bf Amazon datasets}: We use 5 different Amazon category datasets from \cite{he2016ups}: \emph{Apps for Android}, \emph{Cellphone Accessories}, \emph{Pet Supplies}, \emph{Toys and Games}, and \emph{Video Games}. \textcolor{red}{As top-level product categories on Amazon are treated as separate datasets \cite{kang2018self}, the five selected categories are chosen as to promote sparsity, variability, and data size.}

    \item {\bf Amazon datasets \cite{he2016ups}}:  As top-level product categories on Amazon are treated as independent datasets \cite{kang2018self}, we use 5 different Amazon category datasets to vary the sparsity, variability, and data size: \emph{Apps for Android}, \emph{Cellphone Accessories}, \emph{Pet Supplies}, \emph{Toys and Games}, and \emph{Video Games}.

    \item {\bf Yelp dataset}: This is a user rating dataset on businesses. We use the dataset obtained from \cite{he2016fast}.%, where repetitive user-item interactions are merged to the earliest one to avoid training interactions appearing in the test set.
%\end{itemize}
\squishend

For data preprocessing, we adopted a popular \emph{k-core} preprocessing step \cite{he2016ups} (with $k$=5), filtering out users and items with less than 5 interactions. All observed ratings are considered as positive interactions and the remaining as negative interactions. The maximum number of short-term items is set to $s=5$ in all datasets as it covers the short-term peak (see Figure \ref{fig:train_stat}). Table \ref{table:datasets} summarizes the statistics of all datasets, as well as their number of long-term items $l$.
% In long term, we observe that several users in Amazon datasets consumed an extremely large number of items compared to the majority of users. It will cost large amount of unnecessary zero paddings if we decide to cover all these outlier users. %To select the maximum number of long-term items $l$, we observe that several users in Amazon datasets consumed an extremely large number of items compared to the majority of users. If we consider those extreme users, majority of users' historical items will be unnecessarily padded zeroes.
% Hence, we set $l$ to the upper bound of the boxplot approach (i.e. Q3 + 1.5IQR, where Q3 is the \emph{third quartile}, and IQR is the \emph{Interquartile range} of the sequence length distribution of all users). Table \ref{table:datasets} summarizes the statistics of all datasets.

\vspace{-9pt}
\subsection{State-of-the-art Baselines}
We compared our proposed models with \textbf{11} strong state-of-the-art recommendation models as follows:
%\begin{itemize}
\squishlist
    \item {\bf AASVD}: It is an attentive version of the well-known Asymmetric SVD model (ASVD) \cite{koren2008factorization},
    where real-valued self-attention is applied to measure attentive contribution of previously consumed items by a user.

    %In ASVD, users are represented by averaging their consumed items' latent representations equally. AASVD extends ASVD by measuring attentive contribution of consumed items. %Note that AASVD uses the same self-attention with our proposed QAASVD, where Hamilton products and Quaternion embeddings are replaced by dot product and real-valued embeddings.
    \item {\bf QCF} \cite{zhang2019quaternion}: It is a state-of-the-art recommender that represents users/items by Quaternion embeddings.
    \item {\bf NeuMF++} \cite{he2017neural}:
    % It is a state-of-the-art neural recommender that
    It models non-linear user-item interactions by using a MLP and a Generalized MF (GMF) component. We pretrained MLP and GMF to obtain NeuMF's best performance.
    \item {\bf NAIS} \cite{he2018nais}: It is an extension of \emph{ASVD} where contribution of consumed items to the target item is attentively assigned. We adopt $NAIS_{prod}$ version as it led to its best results.
    \item {\bf FPMC} \cite{rendle2010factorizing}: It is a state-of-the-art sequential recommender. It uses the first-order Markov to model the transition between the next item and the previously consumed items.
%    \item {\bf FPMC} \cite{rendle2010factorizing}: a state-of-the-art sequential recommender that model the first-order Markov transition between the next item and the previously consumed items.
    \item {\bf AGRU}: It is an extension of the well-known GRU4Rec \cite{hidasi2015session}, where we use an attention mechanism to combine different hidden states. We experiment with two attention mechanisms: real-valued self-attention, and real-valued \emph{prod attention} proposed by \cite{he2018nais}. Then we report its best performance.
    
    \item {\bf ALSTM}: It is a LSTM based model. Similar to AGRU, we experiment with the real-valued self-attention and the \emph{prod attention} \cite{he2018nais}, and then report its best results.
    % \item {\bf ALSTM}: It is an attentional LSTM based model. Similar to AGRU, we also experiment with the real-valued self-attention and the \emph{prod attention}, and then report its best results.
    
    \item {\bf Caser} \cite{tang2018personalized}:
    % It is a widely used sequential recommender. %It uses convolution neural network design.
    It embedded a sequence of recently consumed items into an ``image'' in time and latent spaces, and uses convolution neural network to learn sequential patterns as local features of an image using different horizontal and vertical filters.
     \item {\bf SASRec} \cite{kang2018self}: It is a strong sequential recommender model. It uses the self-attention mechanism with a multi-head design to identify relevant items for next predictive items.
    \item \textbf{Sli-Rec} \cite{yu2019adaptive}: It uses a time-aware controller to control the state transition. Then it uses an attention-based framework to fuse a user's long-term and short-term preferences.
    \item {\bf ALSTM+AASVD}: 
    It is our implementation that resembles the same architecture as our proposed Quaternion fusion approach, except that it uses Euclidean space instead of Quaternion space. The purpose of implementing and using it as a baseline is to present the effectiveness of our framework and Quaternion representations over the real-valued representations.
    % Though ALSTM + AASVD is not an existing baseline, we implement it as a baseline by resembling the same architecture as our proposed Quaternion fusion approach, except that it uses Euclidean space instead of Quaternion space. Our goal is to present the effectiveness of our framework and Quaternion representations over the real-valued representations in Euclidean space.
    % Though ALSTM + AASVD is not an existing baseline, we implement it as a baseline by resembling the same architecture as our proposed Quaternion fusion approach, except that it uses Euclidean space instead of Quaternion space. 
    %The purpose of implementing and using it as a baseline is to present the effectiveness of our framework and Quaternion representations over the real-valued representations in Euclidean space. The detailed result is described in the following subsection.
    %incorporating estimated
    %learning both long-term and short-term user interests.
%\end{itemize}
\squishend

First four baselines (AASVD, QCF, NeuMF++, and NAIS) are classified as user's long-term interest encoding models. Next four baselines (FPMC, AGRU, ALSTM, and Caser) are user's short-term interest encoding models, and \emph{SASRec, SLi-Rec, and ALSTM+AASVD} encode both user's long-term and short-term intents.
Note that we performed an experiment with \emph{DIEN} \cite{zhou2019deep} (i.e. a long short-term modeling baseline) based on the authors' public source code, which produced surprisingly low results, so we omit its detailed results.
% Note that we also experimented ASVD \cite{koren2008factorization}, LSTM, GRU, Quaternion LSTM \cite{parcollet2018quaternion}, SLIM \cite{ning2011slim}, FISM \cite{kabbur2013fism}, MF-BPR, but do not report their results due to space limitation and their worse results compared to the above baselines.
We also experimented with \emph{ASVD}, LSTM, GRU and Quaternion LSTM but do not report their results due to space limitation and their worse results. Similarly, we omit \emph{BPR} \cite{rendle2012bpr} and \emph{FISM} \cite{kabbur2013fism} results due to their less impressive performance.
%We do not report \emph{BPR} because NeuMF++ outperformed it. Similarly, \emph{NAIS} outperformed \emph{SLIM} \cite{ning2011slim}, \emph{FISM} \cite{kabbur2013fism} so we do not report their results.

% We compare the above baselines with our proposals as follows:
% \squishlist
%     \item {\bf QUALE}: Our proposal to model users' long-term preferences.
%     \item {\bf QUASE (GRU/LSTM)}: Our proposal to model users' short-term preferences using the Quaternion self-attentive GRU/LSTM.
%     \item {\bf QUALSE}: Our fusion to model both user's long-term and short-term preferences.
%     \item {\bf AQUALE}: Our Adversarial QUALE model, learning with our proposed Quaternion adversarial learning.
%     \item {\bf AQUASE}: Our Adversarial QUASE model.
%     \item {\bf AQUALSE}: Our Adversarial fused QUALSE model.

 \vspace{-6pt}
\subsection{Experimental Settings}
\noindent\textbf{Protocol}: We adopt a well-known and practical 70/10/20 splitting proportions to divide each dataset into train/validation (or development)/test sets \cite{tran2018regularizing,liang2018variational}. All user-item interactions are sorted in ascending order in terms of the interaction time. Then, the first 70\% of all interactions are used for training, the next 10\% of all interactions are used for development, and the rest is used for testing. We follow \cite{wang2019neural,xin2019relational} to sample 1,000 unobserved items that the target user has not interacted before, and rank all her positive items with these 1,000 unobserved items for testing models.

\noindent\textbf{Evaluation metrics:} We evaluate the performances of all models by using two well-known metrics: \emph{Hit} Ratio (\emph{HIT@N}), and Normalized Discounted Cumulative Gain (\emph{NDCG@N}). \emph{HIT@N} measures whether all the test items are in the recommended list or not, while \emph{NDCG@N} takes into account the position of the test items, and assigns higher scores if test items are at top-rank positions.

\noindent\textbf{Hyper-parameters Settings:} All models are trained with \emph{Adam} optimizer \cite{kingma2014adam}. A learning rate is chosen from \{0.001, 0.0005\}, and regularization hyperparameters are chosen from \{0, 0.1, 0.001, 0.0001\}. An embedding size $d$ is chosen from \{32, 48, 64, 96, 128\}. Note that for Quaternion embeddings, each component value is a vector of size $\frac{d}{4}$. The number of epochs is 30. The batch size is 256. The number of MLP layers in NeuMF++ is tuned from \{1, 2, 3\}. The number of negative samples per one positive instance is 4 for training models. The settings of \emph{Caser}, \emph{NAIS}, \emph{SASRec} are followed by their reported default settings. In training with \emph{QABPR} loss, the regularization $\lambda_{adv}$ is set to 1. The noise magnitude $\epsilon$ is chosen from \{0.5, 1, 2\}. The adversarial noise is added only in training process, and is initialized as \emph{zero}. All hyper-parameters are tuned by using the validation set.

\vspace{-5pt}
\subsection{Experimental Results}
%Table \ref{table:PerformanceComparison} shows the performances of our models and all the compared baselines on six datasets. We use the results to answer the following research questions:

% \noindent\textbf{RQ1: Our models vs baselines?}
\subsubsection{\textbf{RQ1: Performance comparison}}
Table \ref{table:PerformanceComparison} shows that our proposed fused models \emph{QUALSE}
%(i.e. training with \emph{BPR} loss)
and \emph{AQUALSE}
%(i.e. \emph{QUALSE} but training with our \emph{QABPR} loss)
outperformed all the compared baselines. On average, \emph{QUALSE} improved \emph{Hit@100} by 5.65\% and \emph{NDCG@100} by 8.44\% compared to the best baseline's performances. \emph{AQUALSE} gains additional improvement over \emph{QUALSE}, enhancing \emph{Hit@100} by 7.39\% and \emph{NDCG@100} by 10.83\% on average compared to the best baseline. The improvement of our proposals over the baselines is significant under the Directional Wilcoxon signed-rank test (\emph{p-value} $< 0.015$). We also observed similar results on all six datasets when we measure \emph{Hit@1} and \emph{NDCG@1}. In particular, our \emph{QUALSE} improved \emph{Hit@1} by 6.87\% and \emph{NDCG@1} by 8.71\% on average compared with the best baseline. \emph{AQUALSE} improved \emph{Hit@1} by 8.43\% and \emph{NDCG@1} by 10.27\% on average compared with the best baseline, confirming its consistent effectiveness.

% 	       HIT-1   NDCG-1
%QUALSE	        6.87	     8.71
%AQUALSE	8.43	   10.27

%Even with \emph{topN=1}, we also observe similar results on all six datasets. On average over six datasets, our proposals improved HIT@1 and NDCG@1 by 8.6\% compared to the best baseline.

\vspace{-6pt}
\subsubsection{Varying top-N recommendation list and embedding size:}
To further provide detailed effectiveness of our proposals, we compare QUALSE and AQUALSE models with the top-5 baselines when varying the embedding size from \{32, 48, 64, 96, 128\} and the \emph{top-N} recommendation list from \{10, 20, 50, 100\}. %The results are shown in Figure \ref{fig:sensitivity-topN-embedsize}.

\begin{figure}[t]
    \includegraphics[width=0.42\textwidth]{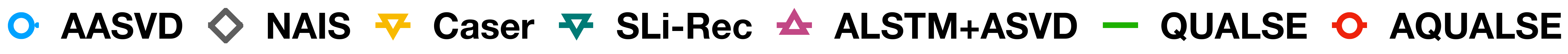}
    \centering
	\begin{subfigure}{0.23\textwidth}
		\centering
		\includegraphics[width=\textwidth]{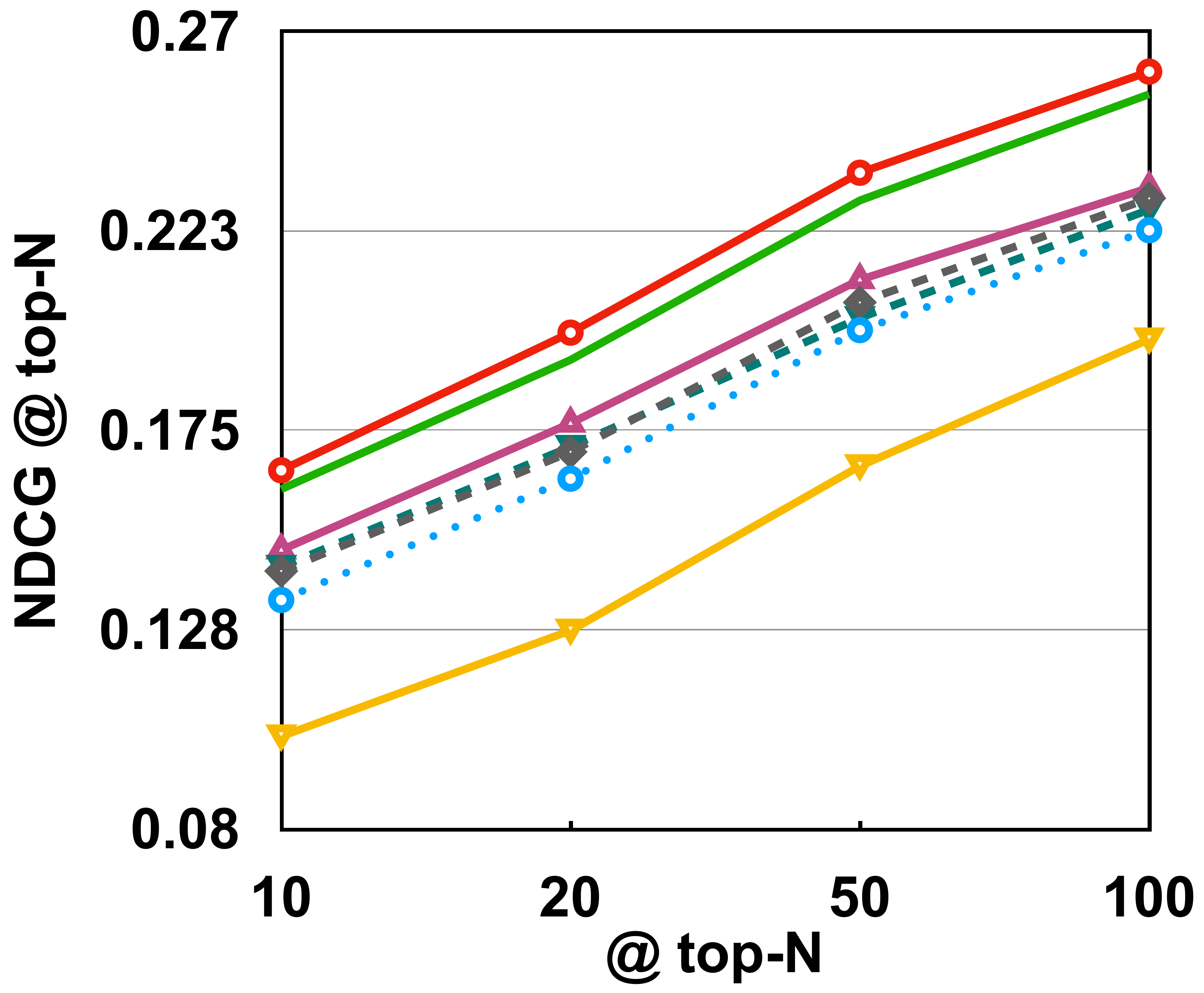}
		\vspace{-15pt}
		\caption{NDCG@topN in Video Games}
		\label{fig:video-game-varyTopN-ndcg}
	\end{subfigure}
	\begin{subfigure}{0.23\textwidth}
		\centering
		\includegraphics[width=\textwidth]{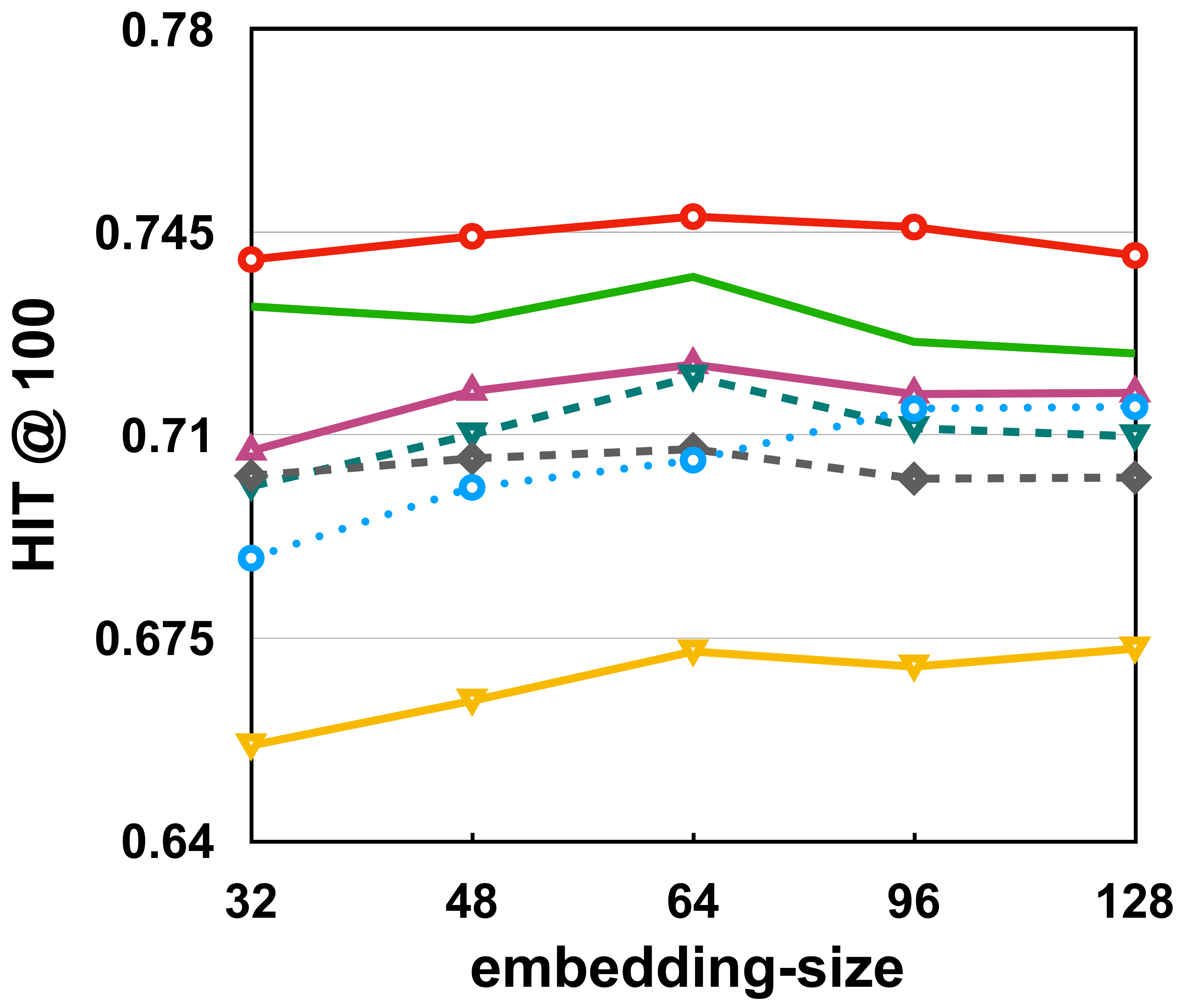}
		\vspace{-15pt}
		\caption{HIT@100 in Yelp.}
		\label{fig:yelp-vary-embed-hit}
	\end{subfigure}
    \vspace{-12pt}
    \caption{Performance of our models and the top-5 baselines when varying a \emph{top-N} recommendation list (left) and an embedding size (right). %Five best baselines are compared with \emph{QUALSE} and \emph{AQUALSE} for a clear visualization.
    }
    \label{fig:sensitivity-topN-embedsize}
    \vspace{-15pt}
\end{figure}

Figure \ref{fig:video-game-varyTopN-ndcg} shows that even with small \emph{top-N} values (e.g., @10), our models consistently outperformed all the compared baselines in the \emph{Video Games} dataset, improving the ranking performance by a large margin of 9.25\%$\sim$12.30\% on average. Specifically, at \emph{top-N}=10 in \emph{Video Games} dataset, QUALSE and AQUALSE improves NDCG@10 over the best baseline by 9.9\% and 12.97\%, respectively.

Figure \ref{fig:yelp-vary-embed-hit} shows the HIT@100 performance of our QUALSE and AQUALSE models, and the top-5 baselines in the Yelp dataset when varying the embedding size. We observe that our proposals outperformed all the baselines. 
Interestingly, while non-adversarial models are more sensitive to the change of the embedding size, our adversarial AQUALSE model is relatively smoother when varying the embedding size. The result makes sense because the adversarial learning reduces the noise effect. Because of the space limitation, we only show detailed results of the \emph{Video Games} and \emph{Yelp} datasets.

% \noindent\textbf{Sensitivity of our Quaternion-based proposals when varying \emph{top-N} recommendation list and embedding size?} We show the NDCG performances of our \emph{QUALSE}, \emph{AQUALSE} and the best five baselines when varying the \emph{top-N} recommendation list from \{10, 20, 50, 100\} (Figure \ref{fig:video-game-varyTopN-ndcg}) on Video Games dataset, and when varying the embedding size from \{32, 48, 64, 96, 128\} (Figure \ref{fig:yelp-vary-embed-ndcg}) in Yelp dataset. We observe that our proposals perform better than the best five baselines. Particularly, on average, \emph{QUALSE} improves \emph{NDCG} by 2.45$\sim$7.52\%, and  \emph{AQUALSE} improves \emph{NDCG} by 4.66$\sim$8.84\% compared to the best baseline performances.

%%%%%%%%%% VARY TOP-N figures
%%%%%%%%%%%%%%%%%%%%%%%%%%%%%%%%%%%%%%%%%%%
% \noindent\textbf{RQ2: Long-term vs short-term vs fused models?}
\vspace{-3pt}
\subsubsection{\textbf{RQ2: Effect of the long-term and short-term encoding components?}}
\begin{figure}[t]
    \centering
   \includegraphics[width=0.8\linewidth]{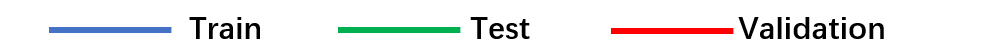}
    \begin{subfigure}{0.23\textwidth}
		\centering
		\includegraphics[width=\textwidth, height=70pt]{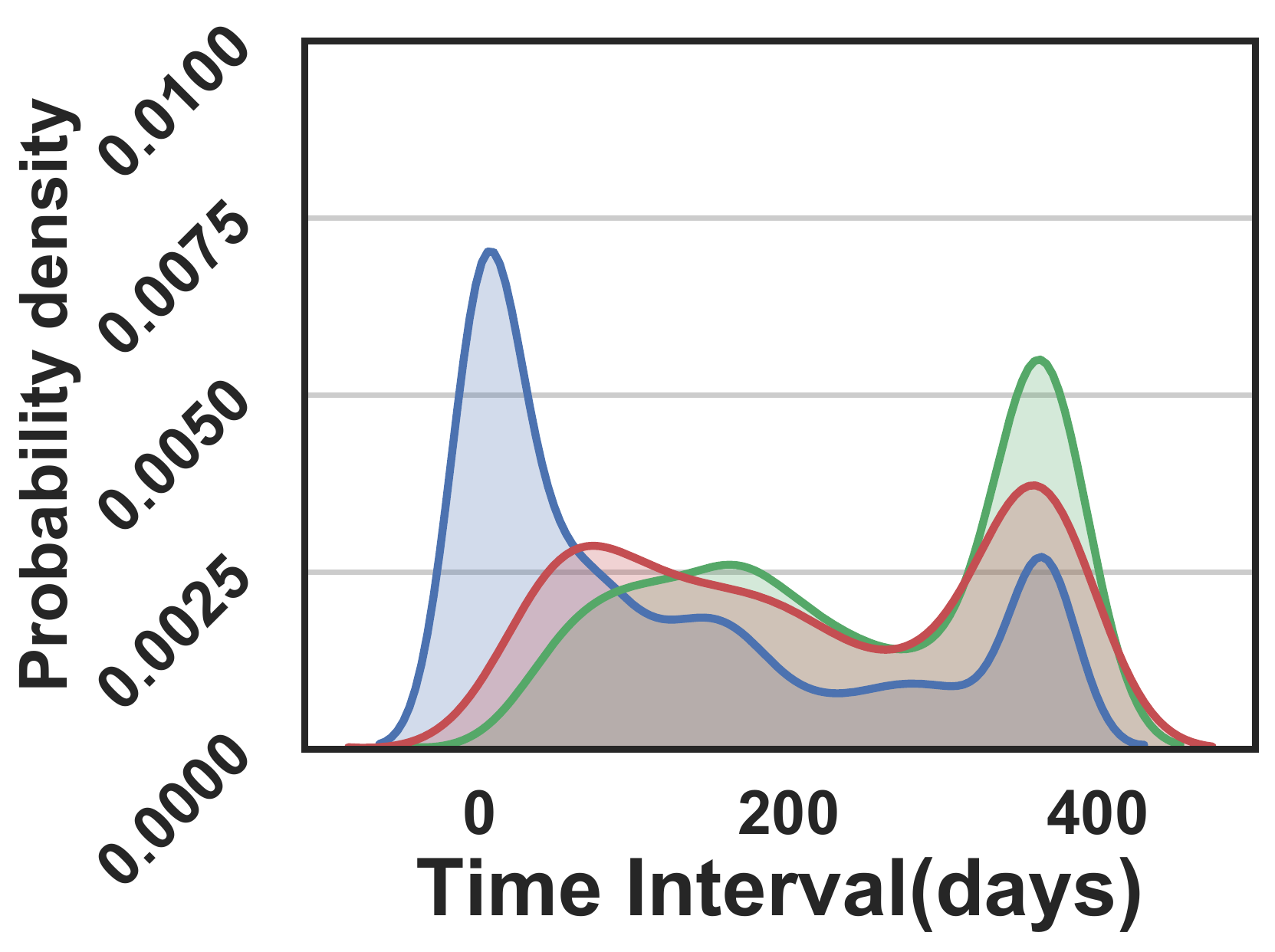}
		\vspace{-10pt}
		\caption{Pet Supplies dataset.}
		\label{fig:density-exp-pet}
	\end{subfigure}
    \begin{subfigure}{0.23\textwidth}
		\centering
		\includegraphics[width=\textwidth, height=70pt]{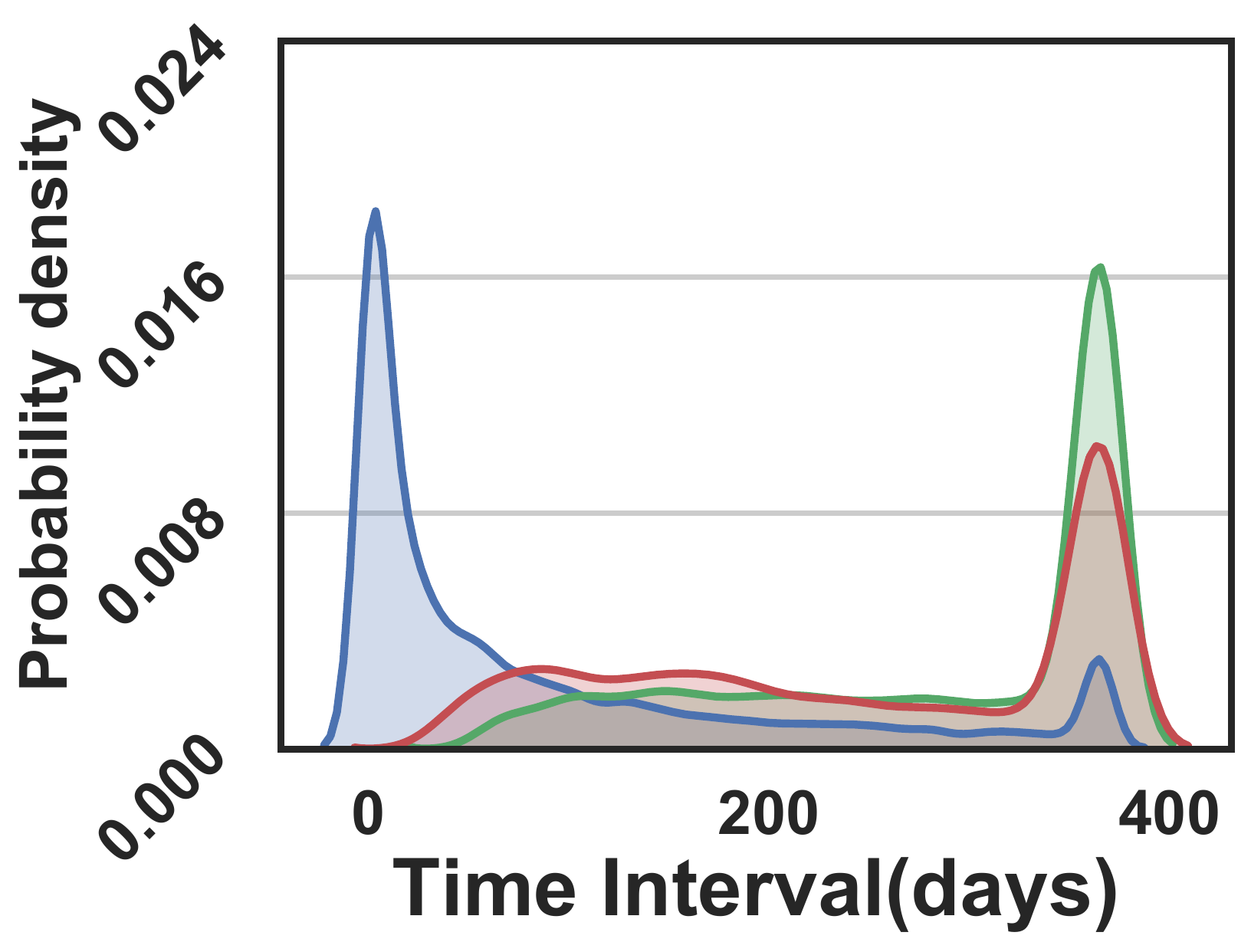}
		\vspace{-10pt}
		\caption{Yelp dataset.}
		\label{fig:density-exp-yelp}
	\end{subfigure}
	\vspace{-10pt}
    \caption{Density distribution of item-item similarity scores in train/vad/test sets of \emph{Pet Supplies} and \emph{Yelp} datasets.}
    \label{fig:density-exp}
    \vspace{-15pt}
\end{figure}

Using reported results in Table \ref{table:PerformanceComparison}, we first compare long-term encoding models (i.e. (a)-(d), and \emph{QUALE}, \emph{AQUALE}) with short-term encoding models (i.e. (e)-(h), and \emph{QUASE}, \emph{AQUASE}). In general, long-term encoding models work better than short-term encoding models. For instance, NAIS (i.e. best long-term encoding baseline) improves 8.5\% on average on six datasets compared with Caser (i.e. best short-term encoding baseline). Similarly, our long-term encoding \emph{QUALE} model works better than our short-term encoding \emph{QUASE} model, enhancing 9.2\% on average over six datasets. To investigate this phenomenon, we plot the density distribution of item-item similarity scores in test sets of two datasets \emph{Pet-Supplies} and \emph{Yelp} in Figure \ref{fig:density-exp}. We observe higher peaks on long-term item-item relationships in the curves, explaining why long-term encoding models work better than short-term encoding models.

Next, we compare the fused models with models that encode either long-term or short-term users preferences. Table \ref{table:PerformanceComparison} shows that models, which consider both user's long-term and short-term preferences, work better than other models, which encode either user's long-term or short-term interests. Both (j) and (k) baselines generally work better than (a)--(h) baselines. Specifically, our QUALSE and AQUALSE models improve 7.9\%$\sim$10.0\% on average over six datasets compared to the best baseline from (a)--(h). These observations show the effectiveness of modeling both user's long-term and short-term interests.
Among models, which consider both user long-term and short-term interests, SASRec performed the worst compared to baselines (i)--(k) and our QUALSE and AQUALSE. This is due to the fact that SASRec models user's long-term and short-term interests implicitly and concurrently by using the Transformer multi-head attention mechanism. But, SLi-Rec, ALSTM+AASVD, and our proposals model the two preferences explicitly and separately, and then combine them later on, increasing flexibility. Note that, although SLi-Rec employed a time-aware attentional LSTM to better model the user's short-term preferences, our ALSTM+AASVD implementation works slightly better than SLi-Rec due to its two distinct properties: (i) the personalized self-attention in AASVD, where each user is parameterized by her own context vector, and (ii) the personalized gating fusion. %To further confirm the effectiveness of these two properties, we experiment without using the personalized self-attention (i.e. use a single global context vector for all users). Performance of AASVD and \emph{QUALE} was dropped by 3.71\% and 2.53\% on average over six datasets, respectively. Similarly, without using the personalized gating fusion, performance of ALSTM+AASVD and \emph{QUALSE} was dropped by 2.63\% and 1.95\% on average over six datasets, respectively.

%AASVD	                QUALE
%-3.71%	                 -2.53%
%AASVD + ALSTM	QUALSE
%-2.63%	                 -1.95%

%%%%%%%%%% DENSITY SCORES
%%%%%%%%%%%%%%%%%%%%%%%%%%%%%%%%%%%%%%%%%%%

\vspace{-4pt}
\subsubsection{\textbf{RQ3: Is using Quaternion representation helpful?}}
In Table \ref{table:PerformanceComparison}, we compare different model pairs: \emph{AASVD} vs. \emph{QUALE}, \emph{ALSTM} vs. \emph{QUASE (LSTM)}, \emph{AGRU} vs. \emph{QUASE (GRU)}, and \emph{ALSTM+AASVD} vs. \emph{QUALSE}. Two methods under the same pair have similar architecture (again, \emph{ALSTM+AASVD} was implemented by us, following our QUALSE architecture to show effectiveness of Quaternion representation). But, the first method of each pair uses real-valued representations and the second method of each pair uses Quaternion representations. %Note that \emph{AASVD} shares exactly the same architecture with \emph{QUALE}, except that \emph{QUALE} uses Quaternion representations, while \emph{AASVD} uses real-valued representations, and so do the other pairs.
Table \ref{table:PerformanceComparison} shows that \emph{QUALE} works better than \emph{AASVD}. %For example, in \emph{Cellphone Accessories} dataset, \emph{QUALE} enhances \emph{AASVD} by 7.47\% of \emph{Hit@100} and 12.20\% of \emph{NDCG@100}. In \emph{Video Games} dataset, \emph{QUALE} improves \emph{AASVD} by 9.38\% of \emph{Hit@100} and 9.58\% of \emph{NDCG@100}.
In six datasets, on average, \emph{QUALE} improves \emph{Hit@100} by 5.60\% and \emph{NDCG@100} by 7.71\% compared to \emph{AASVD}. Similarly, we observe the same patterns from the other three model pairs. %for other comparisons of pairs of models, we observe that in six datasets, on average, \emph{QUASE (LSTM)} improves \emph{ALSTM} by 11.88\% of \emph{Hit@100} and 22.56\% of \emph{NDCG@100}; \emph{QUASE (GRU)} boosts \emph{AGRU} by 15.11\% of \emph{Hit@100} and 24.97\% of \emph{NDCG@100}; \emph{QUALSE} enhances \emph{ALSTM+ASVD} by 6.16\% of \emph{Hit@100} and 8.54\% of \emph{NDCG@100}.
Moreover, when comparing our long-term encoding \emph{QUALE} and \emph{AQUALE} models with other long-term encoding baselines (a)-(e), our models outperformed the baselines, improving \emph{HIT@100} by 5.16\% and 6.55\%, and enhancing \emph{NDCG@100} by 7.11\% and 9.83\%, respectively. %On average six datasets, \emph{QUALE} improves \emph{HIT@100} by 5.16\% and \emph{NDCG@100} by 7.11\% compared to the best long-term baselines (a)-(e) (\emph{p-value} $< 0.05$), while \emph{AQUALE} improves \emph{HIT@100} by 6.55\% and \emph{NDCG@100} by 9.83\% compared to the best long-term baselines (a)-(e) (\emph{p-value} $< 0.015$).
Similarly, our short-term encoding \emph{QUASE} and \emph{AQUASE} using LSTM also work better than other short-term encoding baselines (f)-(h), improving \emph{HIT@100} by 4.75\% and 8.09\%, and enhancing \emph{NDCG@100} by 10.57\% and 15.33\%, respectively.
%We note that performance of \emph{QUASE} when using Quaternion LSTM is slightly better (+0.97\%) than Quaternion GRU, which is consistent to results of real-valued LSTM and real-valued GRU in the literature \cite{cho2014learning}.
All of these results confirm the effectiveness of modeling user's interests by using Quaternion representations over Euclidean representations.

%%%%%%%%%% ATTENTION ANALYSIS %%%%%%%%%%%%%%%%%%%%%%%%%%%%
\begin{figure}[t]
	\centering
	\begin{subfigure}{0.36\textwidth}
		\centering
		\includegraphics[width=0.48\textwidth]{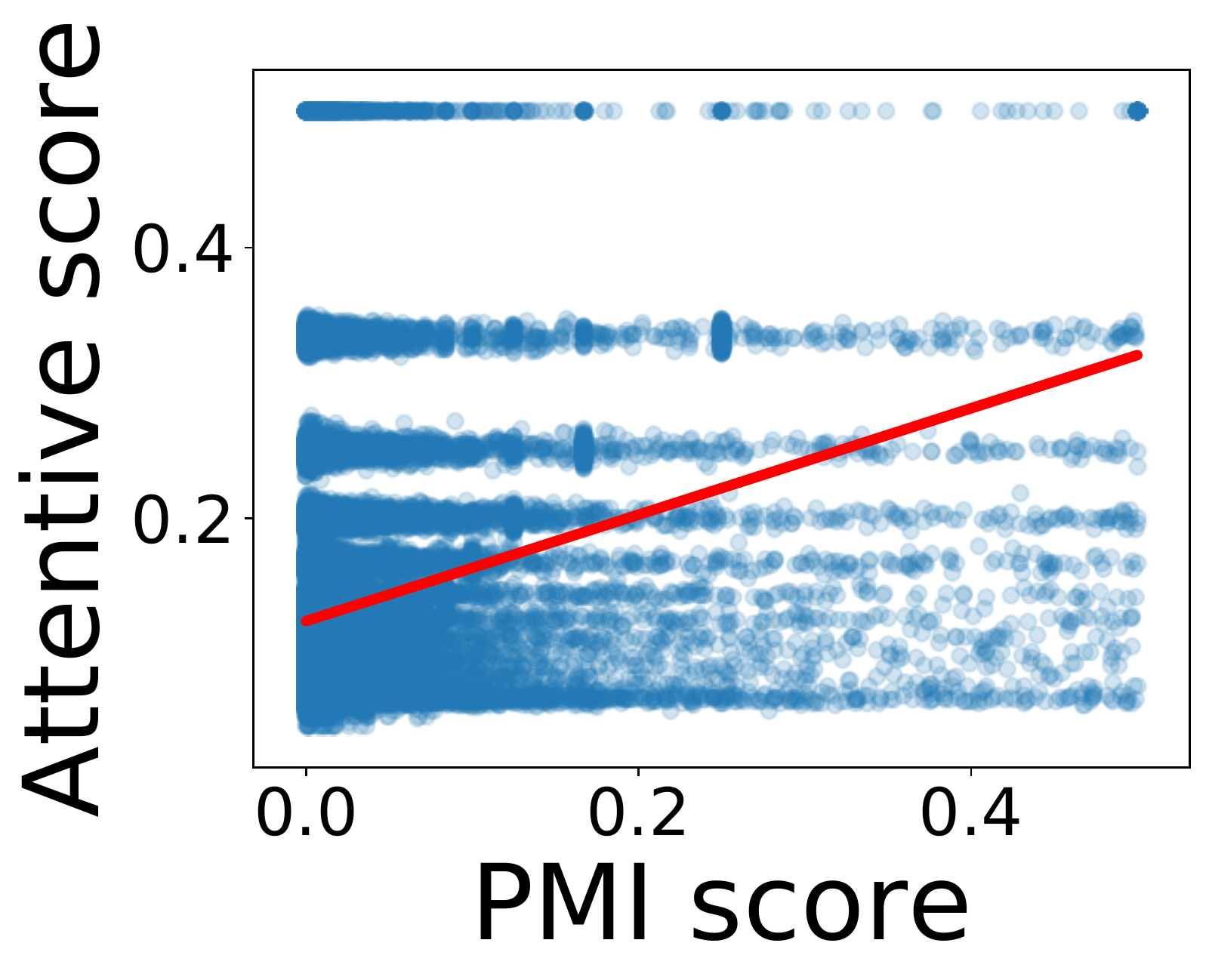}
		\includegraphics[width=0.48\textwidth]{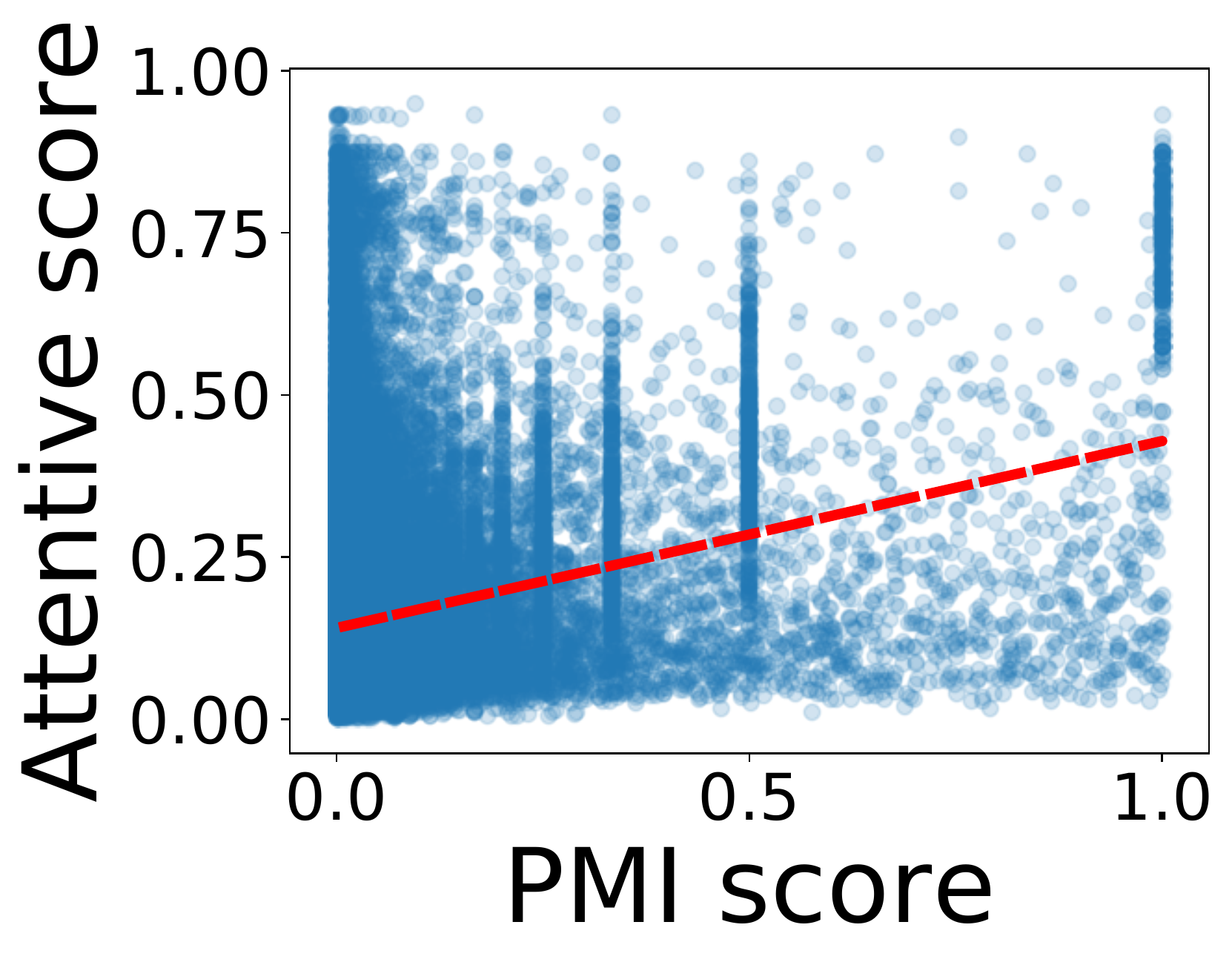}
		\vspace{-5pt}
		% 		\caption{QUALE (left) vs AASVD (right). }%($\rho=0.232$) ($\rho=0.216$).}
		\caption{QUALE vs AASVD. }
		\label{fig:attention-apps-aqsvd}
	\end{subfigure}
	\begin{subfigure}{0.36\textwidth}
		\centering
		\includegraphics[width=0.48\textwidth]{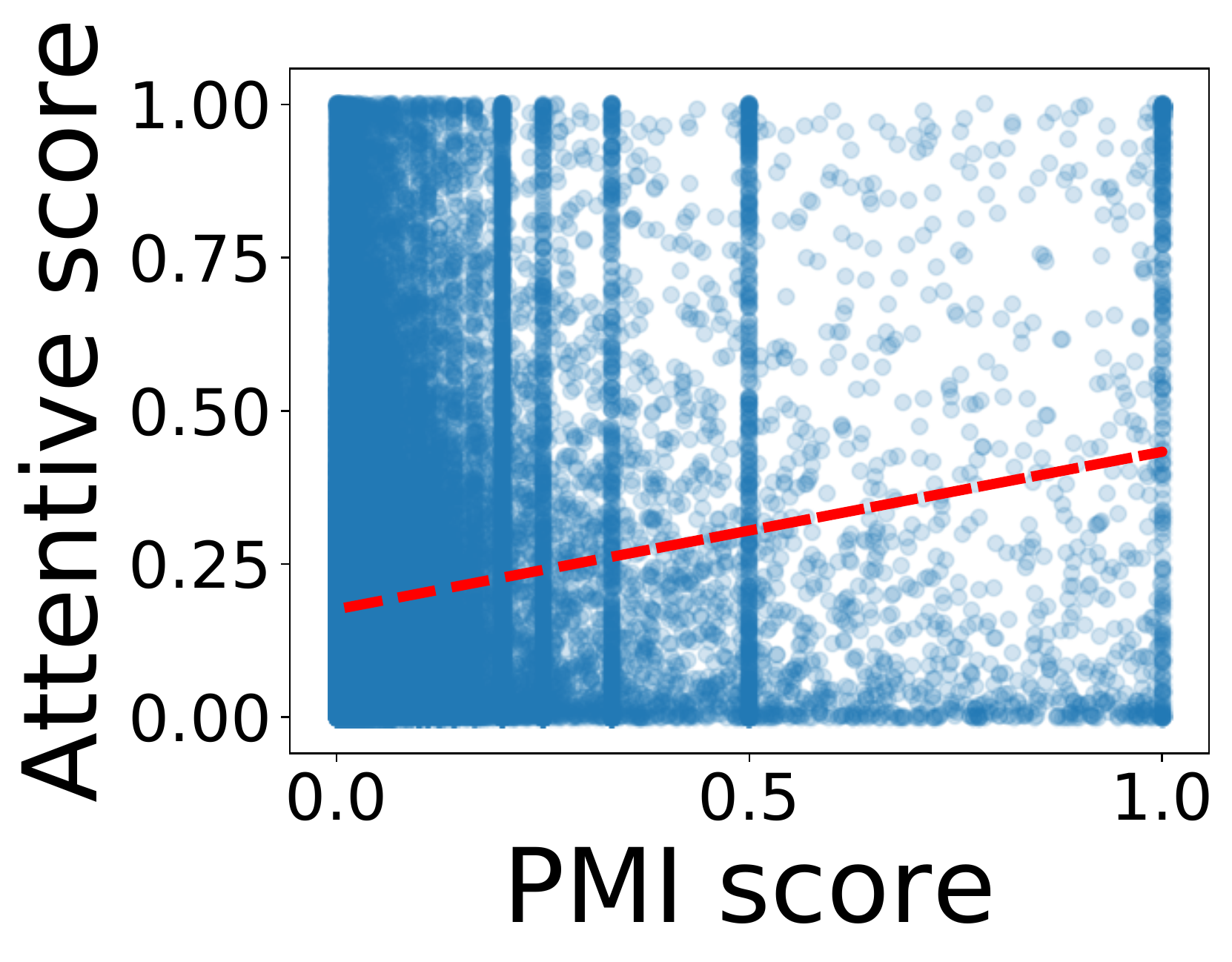}
		\includegraphics[width=0.48\textwidth]{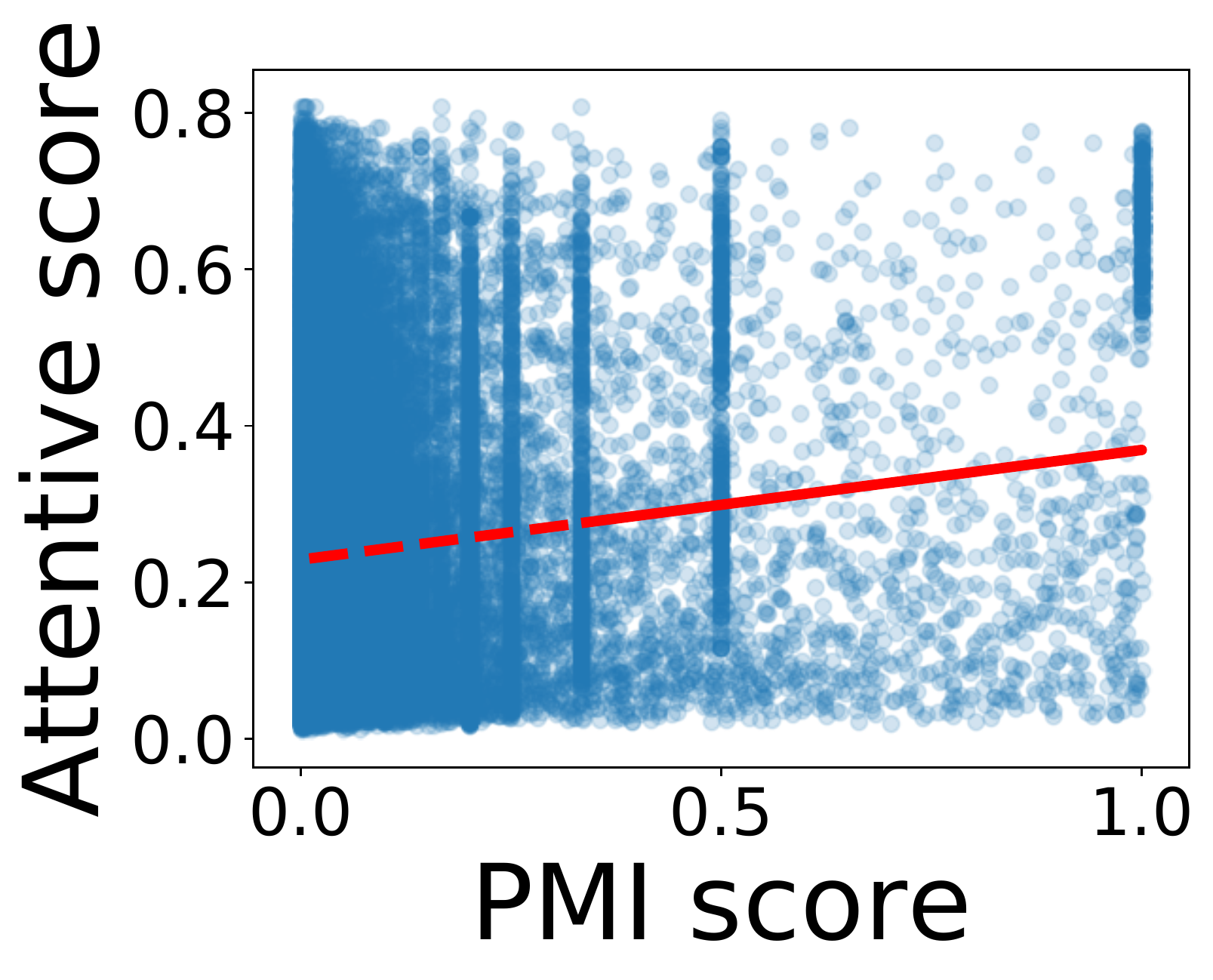}
		\vspace{-5pt}
		% 		\caption{QUASE (left) vs ALSTM (right).}%($\rho=0.148$) ($\rho=0.1$).}
		\caption{QUASE vs ALSTM.}%($\rho=0.148$) ($\rho=0.1$).}
		\label{fig:attention-apps-aqlstm}
	\end{subfigure}
	
	\vspace{-10pt}
	\caption{Comparison of attention scores between (QUALE vs AASVD) and (QUASE vs ALSTM) in the Apps for Android dataset. Pearson correlation $\rho$ between attention scores and PMI scores are: $\rho_{QUALE}=0.232 > \rho_{AASVD}=0.216$, and $\rho_{QUASE}=0.148 > \rho_{ALSTM}=0.1$.}
	%    \caption{Comparison of attention scores in our models (QUALE and QUASE) and in respective real-valued models (AASVD and ALSTM). Pearson correlation $\rho$ between attention scores and PMI scores are: $\rho_{QUALE}=0.232 > \rho_{AASVD}=0.216$, and $\rho_{QUASE}=0.148 > \rho_{ALSTM}=0.1$.}
	\label{fig:attention-analysis}
	\vspace{-10pt}
\end{figure}
%%%%%%%%%%%%%%%%%%%%%%%%%%%%%%%%%%%%%%%%%%%%%%%%%%%%%%%%%%%%

\textbf{Why Quaternion representations help improve the performance?} Since attention mechanism is the key success in deep neural networks \cite{vaswani2017attention}, we analyze how our models assign attention weights compared to their respective real-valued models. We first measure the item-item Pointwise Mutual Information (PMI) scores (i.e. $PMI(j,t) = log \frac{P(j,t)}{P(j)\times P(t)}$) using the training set. The PMI score between two items ($j, t$) gives us the co-occurrence information between item $j$ and item $t$, or how likely the target item $j$ will be preferred by the target user when the item $t$ is already in her consumed item list. We perform \emph{softmax} on all item-item PMI scores. Then, we compare with the generated attention scores from our proposed models and ones from their respective real-valued baseline models. 
Figure \ref{fig:attention-analysis} shows the scatter plots and Pearson correlation comparison using the \emph{Apps for Android} dataset. 
We see that \emph{QUALE, QUASE} tend to correlate more positively with the PMI scores than their respective real-valued models \emph{AASVD, ALSTM}. In another word, our Quaternion-based models assign higher scores for co-occurred item pairs. We reason coming from two aspects of Quaternion representations. First, Hamilton product in Quaternion space encourages strong inter-latent interactions across Quaternion components. Second, since our proposed self-attention mechanism produces scores in Quaternion space, the output attention scores have four values \emph{w.r.t} four Quaternion components. This can be thought as similar to the multi-head attention mechanism \cite{vaswani2017attention} (but not exactly same because of the weight shared in Quaternion transformation), where the proposed attention mechanism learns to attend different aspects from the four Quaternion components. All of these explain why we got better results compared to the respective real-valued models.

\vspace{-4pt}
\subsubsection{\textbf{RQ4: Effect of the personalized gated fusion and the QABPR loss?}}
Table \ref{table:PerformanceComparison} shows that in real-valued representations, \emph{ALSTM+ASVD} works better than \emph{AASVD} and \emph{ALSTM} in all six datasets. %Particularly, in \emph{Toys and Games}, \emph{ALSTM+ASVD} improves \emph{Hit@100} by 1.17\% and \emph{NDCG@100} by 3.04\%  compared to the best performances of \emph{AASVD} and \emph{ALSTM}.
%On average, in six datasets, \emph{ALSTM+ASVD} improves 1.2\% performance compared to the best results of its two degenerated models.
Similarly, in Quaternion representations, the fused \emph{QUALSE} model
%(i.e. fusion of \emph{QUALE} and \emph{QUASE} with LSTM cell)
generally works better than its two degenerated \emph{QUALE} and \emph{QUASE} models. %For instance, \emph{QUALSE} enhances \emph{Hit@100} by 1.41\%  and \emph{NDCG@100} by 3.43\% in \emph{Cellphone Accessories} dataset compared to the best results of its two degenerated models.
% In six datasets, \emph{QUALSE} performs 1.7\% better than the maximum performances of its two degenerated models. %Additionally, \emph{AQUALSE} also works better than \emph{AQUALE} and \emph{AQUASE}, improving \emph{Hit@100} by 1.46\%  and \emph{NDCG@100} by 2.18\% on average compared to the best performances of \emph{AQUALE} and \emph{AQUASE}.
% Additionally, \emph{AQUALSE} also works 1.82\% better than the maximum performances of \emph{AQUALE} and \emph{AQUASE}.
In the six datasets, both \emph{QUALSE} and \emph{AQUALSE} perform better than their degenerated (adversarial) versions, improving 2\% on average \emph{w.r.t} both \emph{HIT@100} and \emph{NDCG@100}. %Additionally, \emph{AQUALSE} also works better than \emph{AQUALE} and \emph{AQUASE}, improving \emph{Hit@100} by 1.46\%  and \emph{NDCG@100} by 2.18\% on average compared to the best performances of \emph{AQUALE} and \emph{AQUASE}.
% Additionally, \emph{AQUALSE} also works 1.82\% better than the maximum performances of \emph{AQUALE} and \emph{AQUASE}.
The results confirm the effectiveness of fusing long-term and short-term user preferences in both of \emph{QUALSE} and \emph{AQUALSE}. %their improvements are not highly boosted. To investigate the reason, we observe the density distribution of item-item similarity scores by using the training, validation and test sets, separately.

We further compare our gating fusion with a weight fixing method, where we vary a contribution score $c \in [0, 1]$ for the user's short-term preference encoding part and $1-c$ for the long-term part. We see that the gating fusion improves 4.82\% on average over six datasets compared to the weight fixing method, again confirming the effectiveness of our personalized gating fusion method. %It makes sense, as the weight fixing method assumes the same contribution rate of the short-term and long-term interests encoding across all the users, which is not ideal as users' interests are diverse.

\noindent\textbf{Is Quaternion Adversarial training on BPR loss helpful?} We compare our proposed models training with \emph{BPR} loss (i.e. \emph{QUALE, QUASE (LSTM),} and \emph{QUALSE} models) and our proposed models training with \emph{QABPR} loss (i.e. \emph{AQUALE}, \emph{AQUASE (LSTM)}, and \emph{AQUALSE}). First, we observe that \emph{AQUASE} boosted \emph{QUASE} performance by a large margin: improving \emph{HIT@100} by 3.2\% and \emph{NDCG@100} by 4.29\% on average in the six datasets. \emph{AQUALE} and \emph{AQUALSE} also improve \emph{QUALE} and \emph{QUALSE} by 1.92\% and 1.91\% on average of both \emph{HIT@100} and \emph{NDCG@100} over six datasets, respectively. These results show the effectiveness of the adversarial attack on Quaternion representations with our \emph{QABPR} loss. %We note that the improvement on \emph{NDCG} ranking measurement is higher than \emph{HIT} when using \emph{AQBPR} loss is mostly due to the nature of \emph{BPR} loss which aims to ameliorate ranking scores. Additionally, though being effective, \emph{QUALE} is much more simpler than \emph{QUASE} model, leading to a more robust model and a smaller improvement of \emph{AQUALE} on \emph{QUALE}, compared to the enhancement of \emph{AQUASE} on \emph{QUASE}.

%% file: 6-conclusion.tex
\vspace{-5pt}
\section{Conclusion}
In this paper, we have shown that user's short-term and long-term interests are complementary and both of them are indispensable. We fully utilized Quaternion space and proposed three novel Quaternion-based models: (1) a \emph{QUALE} model learned the user's long-term intents, (2) a \emph{QUASE} model learned the user's short-term interests, and (3) a \emph{QUALSE} model fused \emph{QUALE} and \emph{QUASE} to learn both user's long-term and short-term preferences. We also proposed a Quaternion-based Adversarial attack on Bayesian Personalized Ranking (QABPR) loss to improve the robustness of our proposals. Through extensive experiments on six real-world datasets, we showed that our \emph{QUALSE} improved 6.87\% at \emph{HIT@1} and 8.71\% at \emph{NDCG@1}, and \emph{AQUALSE} improved 8.43\% at \emph{HIT@1} and 10.27\% at \emph{NDCG@1} on average compared with the best baseline. Our proposed models consistently achieved the best results when varying \emph{top-N} (e.g., HIT@100 and NDCG@100). These results show the effectiveness of our proposed framework.
%In this paper, we propose novel Quaternion-based recommender system, namely QUaternion-based self-Attentive Long Short term user Encoding(QUALSE) model.

%We argue that users' short-term interest and long-term interest are complementary and both of them are indispensable. Concurrent effect of both interest helps to generate better user representation and thus improve recommendation performance. We further extend the personalized model into hypercomplex-valued network by exploring quaternions. We investigated and clarified the benefits of quaternion-based neural network in terms of both mathematics and experimental results. Extensive experiments show our proposed model outperforms strong state-of-the-art baselines. Moreover, each components proves to effectively contribute to our final recommendation. We regard this work as an important step forward for a light-weight and more generalized neural network on sparse data. In the future, we will extend quaternion spaces to graph convolution neural networks for further improve the models in sparse datasets.
%        	       HIT-1   NDCG-1
%QUALSE	        6.87	     8.71
%AQUALSE	8.43	   10.27

\vspace{-5pt}
\section{ACKNOWLEDGMENTS}
This work was supported in part by NSF grant CNS-1755536, AWS Cloud Credits for Research, and Google Cloud. 
% Any opinions, findings and conclusions or recommendations expressed in this material are the author(s) and do not necessarily reflect those of the sponsors.